\documentclass{aa}

\usepackage{multirow}
\usepackage{amsmath}
\usepackage{amssymb}
\usepackage{graphicx}
\usepackage{fancybox}
\usepackage{psfrag}
\usepackage{pstricks}
\usepackage{epsf}
\usepackage{epsfig}
\usepackage{txfonts} 
\usepackage{natbib}

\def\eg{\mbox{e.g.~}}
\def\ie{\mbox{i.e.~}}
\def\Rin{R_\mathrm{in}}
\def\Tin{T_\mathrm{in}}
\def\Rout{R_\mathrm{out}}
\def\RT{R_\mathrm{T}}
\def\microns{\mu\mathrm{m}}
\def\kappaabs{\kappa^\mathrm{abs}_\lambda}
\def\kappaext{\kappa^\mathrm{ext}_\lambda}
\def\rr{\vec{r}}
\def\Cabs{C^\mathrm{abs}_\lambda}
\def\Cext{C^\mathrm{ext}_\lambda}

\def\dMdO{\frac{d\dot{M}}{d\Omega}}
\def\rhoin{n_\mathrm{in}}

\def\dthetadisc{\Delta\theta_\mathrm{d}}

\def\hatn{\hat{n}}
\def\hatx{\hat{x}}
\def\haty{\hat{y}}
\def\hatz{\hat{z}}
\def\hatX{\hat{X}}
\def\hatY{\hat{Y}}
\def\etal{\eta_\lambda}
\def\Bl{B_\lambda}
\def\Il{I_\lambda}
\def\Isl{I^{\mathrm{s}}_\lambda}
\def\ss{s^\mathrm{(s)}}
\def\s{\mathrm{s}}
\def\e{\mathrm{e}}
\def\Rs{R_\mathrm{s}}

\def\phis{I^{\mathrm{s}}_{\lambda_0}}
\def\taul{\tau_\lambda}

\def\Vcell{V_\xi}
\def\zcell{z_\xi}
\def\rhocell{\rho_\xi}
\def\Deltacell{\Delta_\xi}
\def\Vtot{V_\mathrm{tot}}

\def\PAd{\mathrm{PA}_\mathrm{d}}
\def\Vl{V_{\lambda}}
\def\PA{\mathrm{PA}}
\def\smax{s_\mathrm{max}}

\def\ncells{n_\mathrm{cells}}
\def\taull{\tau_{\lambda}^{(l)}}

\def\amin{a_\mathrm{min}}
\def\amax{a_\mathrm{max}}
\def\W{\mathrm{W}}
\def\m{\mathrm{m}}
\def\Rsun{R_{\sun}}
\def\m{\mathrm{m}}
\def\K{\mathrm{K}}
\def\W{\mathrm{W}}
\def\str{\mathrm{str}}
\def\msunyr{M_{\sun}\,\mathrm{yr^{-1}}}
\def\nlambda{n_\lambda}
\def\Vobs{V^\mathrm{obs}}
\def\Fl{F_\lambda}
\def\Fobs{F^\mathrm{obs}}
\def\sigmaV{\sigma_{V}}
\def\sigmaF{\sigma_{F}}

\def\nbase{n_\mathrm{B}}
\def\chisV{\chi^{2}_{|V|}}
\def\chisF{\chi^{2}_{F}}
\def\H{\mathrm{H}}
\def\k{\mathrm{k}}
\def\pc{\mathrm{pc}}
\def\B{\mathrm{B}}
\def\G{\mathrm{G}}
\def\Hz{\mathrm{Hz}}

\def\Deltax{\Delta_{X}}
\def\Deltay{\Delta_{Y}}
\def\chir{\chi_\mathrm{r}}
\def\chirmin{\chi_\mathrm{r,min}}

\def\taulim{\Delta \tau_\mathrm{lim}}
\def\Teff{T_\mathrm{eff}}

\begin{document}







\title{Fast ray-tracing algorithm for circumstellar structures
  (FRACS)}

\subtitle{I. Algorithm description and parameter-space study for
  mid-IR interferometry of B[e] stars}

\author{G. Niccolini
  \and
  P. Bendjoya
  \and
  A. Domiciano de Souza
}

\institute{
  UMR 6525 CNRS H. FIZEAU-Universit\'e de Nice Sophia-Antipolis,
  Observatoire de la C\^{o}te d'Azur  Campus Valrose,
  F-06108 Nice cedex 2, France
}



\abstract
{}
{The physical interpretation of spectro-interferometric data is
  strongly model-dependent. On one hand, models involving elaborate
  radiative transfer solvers are too time consuming in general to
  perform an automatic fitting procedure and derive astrophysical
  quantities and their related errors. On the other hand, using simple
  geometrical models does not give sufficient insights into the
  physics of the object. We propose to stand in between these two
  extreme approaches by using a physical but still simple
  parameterised model for the object under consideration. Based on
  this philosophy, we developed a numerical tool optimised for
  mid-infrared (mid-IR) interferometry, the fast ray-tracing algorithm
  for circumstellar structures (FRACS) which can be used as a
  stand-alone model, or as an aid for a more advanced physical
  description or even for elaborating observation strategies.}
{FRACS is based on the ray-tracing technique without scattering, but
  supplemented with the use of quadtree meshes and the full symmetries
  of the axisymmetrical problem to significantly decrease the
  necessary computing time to obtain e.g. monochromatic images and
  visibilities. We applied FRACS in a theoretical study of the dusty
  circumstellar environments (CSEs) of B[e] supergiants (sgB[e]) in
  order to determine which information (physical parameters) can be
  retrieved from present mid-IR interferometry (flux and visibility).
}
{From a set of selected dusty CSE models typical of sgB[e] stars we
  show that together with the geometrical parameters (position angle,
  inclination, inner radius), the temperature structure (inner dust
  temperature and gradient) can be well constrained by the mid-IR data
  alone. Our results also indicate that the determination of the
  parameters characterising the CSE density structure is more
  challenging but, in some cases, upper limits as well as correlations
  on the parameters characterising the mass loss can be obtained. Good
  constraints for the sgB[e] central continuum emission (central star
  and inner gas emissions) can be obtained whenever its contribution
  to the total mid-IR flux is only as high as a few percents.
  Ray-tracing parameterised models such as FRACS are thus well adapted
  to prepare and/or interpret long wavelengths (from mid-IR to radio)
  observations at present (e.g. VLTI/MIDI) and near-future (e.g.
  VLTI/MATISSE, ALMA) interferometers.}
{}



\maketitle

\section{Introduction}




When dealing with optical/IR interferometric data, one needs to invoke
a model for the understanding of the astrophysical object under
consideration. This is because of (1) the low coverage of the uv-plane
and most of the time because of the lack of the visibility phase, and
(2) because our aim is to extract physical parameters from the
data. This is particularly true for the Mid-Infrared Interferometric
Instrument~\citep[MIDI, ][]{leinert2003} at the Very Large Telescope
Interferometer (VLTI), on which our considerations will be
focused. Some pure geometrical information can be recovered through a
simple toy model such as Gaussians~\citep[see
\eg][]{leinert2004,domiciano2007}.


However, this approach does not give any insights into the physical
nature of the object. One would dream of having a fully consistent
model to characterise the object under inspection. In many cases, if
not all, a fully consistent model is out of reach and one uses at
least a consistent treatment of the radiative transfer. Models based
for instance on the Monte Carlo method are very popular~\citep[see
\eg][]{ohnaka2006,niccolini2006,wolf1999} for this purpose. Still, the
medium density needs to be parameterised and it is not determined in a
self-consistent way. For massive stars for instance, it would be
necessary to take into account non-LTE effects including both gas and
dust emission of the circumstellar material as well as a full
treatment of radiation hydrodynamics. Fitting interferometric data
this way is as yet impossible because of computing time limitations.


Of course, solving at least the radiative transfer in a
self-consistent way is already very demanding for the computational
resources. Consequently, model parameters cannot be determined in a
fully automatic way and the model fitting process must be carried out
mostly {\sl by hand}, or automatised by systematically exploring the
parameter-space~, the ``chi-by-eye'' approach mentioned
in~\citet{press1992}. The followers of this approach consider the
``best fitting'' model as their best attempt: a model that is
compatible with the data. It is admittedly not perfect, but it is in
most of the cases the best that can be done given the difficulty of
the task. It is remarkable that a thorough $\chi^2$ analysis of
VLTI/MIDI data of the Herbig Ae star AB Aurigae has been performed
by~\citet{difolco2009} which remains to date one of the most achieved
studies of this kind. From the $\chi^2$ analysis, formal errors can be
derived and at least the information concerning the constraints for
the physical parameters can be quantified. Qualitative information
about the correlation of parameters can be pointed out.


The next step after the toy models for the physical characterisation
of the astrophysical objects can be made from the pure geometrical
model towards the self-consistency by including and parameterising the
object emissivity in the analysis. For instance, \citet{lachaume2007}
and~\citet{malbet2005} use optically thick (\ie emitting as black
bodies) and infinitely thin discs to model the circumstellar
environment of pre-main-sequence and B[e] stars. Of course this
approach has some restriction when modelling a disc: for instance it
cannot handle nearly edge-on disc and an optically thin situation.






We propose an intermediate approach: between the use of simple
geometrical models and sophisticated radiative transfer solvers.
Indeed, it is a step backwards from the ``self-consistent'' radiative
transfer treatment, which is in most cases too advanced with regard to
the information provided by the interferometric data. For this
intermediary approach, we assume a prescribed and parameterised
emissivity for the medium. Our purpose is to derive the physical
parameters that characterise this emissivity. In the process, we
compute intensity maps and most particularly visibility curves from
the knowledge of the medium emissivity with a fast ray-tracing
technique (a few seconds depending on map resolution), taking into
account the particular symmetries of a disc configuration. Then, the
model fitting process can be undertaken in an automatic way with
standard methods~\citep[see \eg][]{levenberg1944,marquardt1963}. The
techniques we present are designed to be quite general and not
tailored to any particular emissivity except for the assumed
axisymmetry of the problem under consideration.

Our purpose is twofold. On one hand - as already mentioned - we aim to
estimate physical parameters and their errors characterising the
circumstellar dusty medium under consideration with as few restrictive
assumptions as possible; at least within the obvious limitations of
the present model. On the other hand our purpose is to provide the
user of a more detailed model, such as a Monte Carlo radiative
transfer code, with a first characterisation of the circumstellar
matter to start with.



In Sect.~\ref{sect:raytracing} we describe the general framework of
the proposed ray tracing technique. In particular how to derive the
observable from the astrophysical object emissivity. In
Sect.~\ref{sect:numimp} we describe the numerical aspects that are
specific to the present ray-tracing technique. In particular, the use
of a quadtree mesh and the symmetries that allow us to speed up the
computation are detailed. In Sect.~\ref{sect:testcase} we focus our
attention on the circumstellar disc of B[e] stars and describe a
parametric model of the circumstellar environment. In
Sect.~\ref{sect:numtests} we analyse artificial interferometric data
generated both from the parametric model itself and from a Monte Carlo
radiative transfer code~\citep{niccolini2006}. Our purpose is not to
fit any particular object, but to present our guideline to the
following question: which physical information can we get from the
data ? A discussion of our results and the conclusions of our work are
given in Sects~\ref{sect:discussion} and~\ref{sect:conclusion}
respectively.



\section{The ray-tracing technique}
\label{sect:raytracing}

We describe here the FRACS algorithm, developed to study stars with
CSEs from mid-IR interferometric observables (\eg visibilities,
fluxes, closure phases). Although FRACS could be extended to
investigate any 3D CSE structures, we focus here on the particular
case of axisymmetrical dusty CSEs. This study is motivated by the
typical data one can obtain from disc-like CSE observed with MIDI, the
mid-IR 2-telescope beam-combiner instrument of ESO's
VLTI~\citep{leinert2003}.


\subsection{Intensity map}


Intensity maps of the object are the primary outputs of the model that
we need to compute the visibilities and fluxes that are directly
compared to the observations. For this purpose, we integrate the
radiation transfer equation along a set of rays (ray-tracing
technique) making use of the symmetries of the problem (see
Sect.~\ref{sect:numimp} for details).

\begin{figure}[!t]
  \centering
  \psfrag{x}{$x$}
  \psfrag{y}{$y$}
  \psfrag{z}{$z$}
  \psfrag{X}{$X$}
  \psfrag{Y}{$Y$}
  \psfrag{XP}{$X'=\text{North}$}
  \psfrag{YP}{$Y'=\text{East}$}
  \psfrag{i}{$i$}
  \psfrag{PA}{$\PAd$}
  \psfrag{plane}{fictitious image plane}
  \psfrag{nhat}{$\hatn$}
  \psfrag{line}{line of sight}
  \psfrag{Xh}{$\hatX$}
  \psfrag{Yh}{$\hatY$}
  \resizebox{\hsize}{!}{\includegraphics{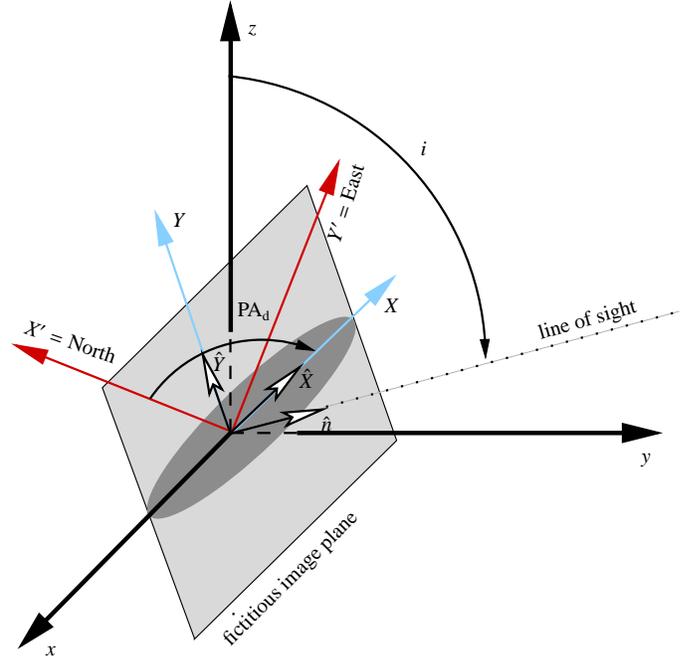}}
  \caption{Coordinate systems. The shaded ellipse represents a disc
    viewed by the observer.}
  \label{fig:geometry}
\end{figure}

The unit vector along the line of sight is given by
$\hatn=\haty\,\sin{i}+\hatz\,\cos{i}$, $i$ being the inclination
between the $z$-axis and the line of sight and $\hatx$, $\haty$ the
unit vector along the $x$ et $y$-axis of a cartesian system of
coordinates (see Fig.~\ref{fig:geometry}), referred to as the ``model
system'' below. The problem is assumed to be invariant by rotation
around the $z$-axis. We define a fictitious image plane by giving two
unit vectors $\hatY=-\haty\,\cos{i}+\hatz\,\sin{i}$ and
$\hatX=-\hatx$. This particular choice is made making use of the
axisymmetry of the problem. Note that for this particular coordinate
system $(X,Y)$ the disc position angle (whenever $i \ne 0$) is {\em
  always} defined as $90\,\degr$. The actual image plane, with the
$Y'$ and $X'$ axis corresponding respectively to North and East, is
obtained by rotating the axis of our fictitious image plane by an
angle $\PAd-\frac{\pi}{2}$, where $\PAd$ is the position angle of the
disc with respect to North.

The dust thermal emissivity at wavelength $\lambda$ and position
vector $\rr$ is given by
\begin{equation}
  \label{eq:emissivity}
  \etal(\rr) = \kappaabs(\rr)\,\Bl(T(\rr)) \ ,
\end{equation}
where $\kappaabs(\rr)$ is the absorption coefficient and $\Bl(T(\rr)$
the Planck function at the medium temperature $T(\rr)$ at $\rr$.
$\kappaabs$ is defined as $n(\rr)\,\Cabs$, where $\Cabs$ is the
absorption cross section and $n(\rr)$ the number density of dust
grains at $\rr$ .

We neglect the scattering of the radiation by dust grains, optimising
our approach to long wavelengths (from mid-IR to radio). This
assumption simplifies the radiative transfer equation by removing the
scattering term.


We obtain the intensity map at position $(X,Y)$ in the image plane
(inclined by $i$) and at wavelength $\lambda$ by integrating the
transfer equation along the particular ray that passes through the
considered point of the image plane. Defining $\rr_s(X,Y,i)$ (simply
$\rr_s$ for short) as the position vector along a ray, given in the
model system of coordinates by
\begin{equation}
  \label{eq:ray}
  \rr_s(X,Y,i)=\left(
    \begin{array}{c}
      -X \\
      -Y\,\cos{i}+s\,\sin{i} \\
      Y\,\sin{i}+s\,\cos{i}
    \end{array} \right) \ ,
\end{equation}
and by introducing the optical depth at wavelength $\lambda$ and
position $s$ along the ray by
\begin{equation}
  \label{eq:tau}
  \taul(X,Y,i;s)=
  \int\limits_{s}^{\sqrt{\Rout^2-R^2}}\!\!\!\!\!\!\kappaext(\rr_{s'})\,ds' \ ,
\end{equation}
we obtain
\begin{equation}
  \label{eq:intensity}
  \Il(X,Y,i)=
  \int\limits_{-\sqrt{\Rout^2-R^2}}^{\sqrt{\Rout^2-R^2}}\!\!\!\!\!\!
  \kappaabs(\rr_s)\,\Bl\left[T(\rr_s)\right]\,
  \e^{\displaystyle{-\taul(X,Y,i;s)}}\,ds \ ,
\end{equation}
where the extinction coefficient $\kappaext(\rr)\approx
\kappaabs(\rr)$ because scattering is neglected.

We assume that the CSE is confined within a sphere of radius $\Rout$,
$s$ varies consequently from $-\sqrt{\Rout^2-R^2}$ to
$\sqrt{\Rout^2-R^2}$ ($R^2=X^2+Y^2$) in Eq.~(\ref{eq:intensity}) and
in the definition of a ray Eq.~(\ref{eq:ray}). This hypothesis can be
relaxed without altering the present considerations and the domain of
integration of Eq.~(\ref{eq:intensity}) suitably chosen.

If some radiation sources (\eg black body spheres) are included in the
analysis, an additional term must be added in Eq.~(\ref{eq:intensity})
whenever a particular ray intersect a source. For a source with
specific intensity $\Isl$ this additional term is given by
$\Isl\,\e^{-\taul(X,Y,i;\ss)}$, $\ss$ being the distance at which the
ray given by $X$, $Y$ and $i$ (see Eq.~\ref{eq:ray}) intersects the
outermost (along the ray) source boundary. In that case the lower
integration limit in Eq.~(\ref{eq:intensity}), that is
$-\sqrt{\Rout^2-R^2}$, must also be replaced by $\ss$.



\subsection{Interferometric observables}

From the monochromatic intensity maps at wavelength $\lambda$
(Eq.~\ref{eq:intensity}) we obtain both the observed fluxes $\Fl$ and
visibilities $\Vl$ for an object at distance $d$,
\begin{equation}
  \label{eq:flux}
\Fl(i)=
\frac{1}{d^2}{\int\limits_{-\infty}^{\infty}\!\int\limits_{-\infty}^{\infty}\, 
  \Il(X,Y,i)\,dXdY} \ ,
\end{equation}
and
\begin{equation}
  \label{eq:visibility}
  \Vl(\B,\PA)=\frac{1}{d^2\,F_\lambda(i)}
   \int\limits_{-\infty}^{\infty}\!\int\limits_{-\infty}^{\infty}\!
    \Il(X,Y,i)\,
    \e^{-2j\pi\,\frac{B}{\lambda}
      \,\left[\frac{X}{d}\cos(\Delta)+\frac{Y}{d}\sin(\Delta)\right]}\,dXdY \ ,
\end{equation}
where $\Vl$ is obtained for a given baseline specified by its
projected length $\B$ (on the sky, \ie $(X',Y')$ coordinates) and its
polar angle $\PA$ from North to East (direction of the $Y'$ axis).
$\Delta$ and $j$ represent, respectively, $\PAd-\PA$ and $\sqrt{-1}$.

\section{Numerical considerations}
\label{sect:numimp}

We seek to produce intensity maps within seconds~\footnote{The actual
  computation time reached is less than $10\,\s$ for a $10^4$ pixel
  map on an Intel T2400 $1.83\,\G\Hz$ CPU.} and we aim for our
numerical method to be sufficiently general in order to deal with a
large range of density and temperature structures. Given these two
relatively tight constraints, the numerical integration of
Eq.~(\ref{eq:intensity}) is not straightforward.

For example we have tested that the $5^\text{th}$ order Runge-Kutta
integrators of~\citet{press1992} with adaptive step-size~\citep[as
discussed in][]{steinacker2006} doest not suit our constraints.
Indeed, the step adaption leads to difficulties if sharp edges (\eg
inner cavities) are present in the medium emissivity.

\subsection{Mesh generation}

Regarding the above mentioned constraints and the different numerical
approaches tested, we found that Eq.~(\ref{eq:intensity}) is more
efficiently computed with an adaptive mesh based on a tree data
structure (quadtrees/octrees). The mesh purpose is twofold: first, it
must guide the computation of Eq.~(\ref{eq:intensity}) and distribute
the integration points along the rays according to the variations of
the medium emissivity; second, within the restriction of
axis-symmetrical situations, the mesh must handle any kind of
emissivity. Quad/octree meshes are extensively used in Monte Carlo
radiative transfer codes~\citep[\eg see][]{bianchi2008, niccolini2006,
  jonsson2006, wolf1999}; the mesh generation algorithm is thoroughly
described in~\citet{kurosawa2001}.


The mesh we use is a {\em cartesian} quadtree. {\em Cartesian} refers
here to the mesh type and not to the system of coordinates we use.
Indeed, the mesh is implemented as a nested squared domain (cells) in
the $\rho-|z|$ plane ($\rho=\sqrt{x^2+y^2}$). The whole mesh is
enclosed by the largest cell (the root cell in the tree hierarchy) of
size $\Rout$ in $\rho$ and $|z|$. The underlying {\em physical}
coordinate system is cylindrical (with $z>0$) and the mesh cells
correspond to a set of two (for $z>0$ and $z<0$) tori, which are the
actual {\em physical} volumes.

The mesh generation algorithm consists in recursively dividing each
cell in four child cells until the following conditions are
simultaneously fulfilled for each cell in the mesh~\citep[see][for
more details]{kurosawa2001}:
\begin{eqnarray}
  \label{eq:mesh1}
  \frac{\iiint\limits_{\Vcell} \left[\kappaabs(\rr)\right]^\alpha \, d^3\rr}
  {\iiint\limits_{\Vtot} \left[\kappaabs(\rr)\right]^\alpha \, d^3\rr} &<& \eta
  \quad \quad \quad \text{and} \\
  \label{eq:mesh2}
  \frac{\iiint\limits_{\Vcell} \left[T(\rr)\right]^\beta \, d^3\rr}
  {\iiint\limits_{\Vtot} \left[T(\rr)\right]^\beta \, d^3\rr} &<& \eta \ ,
\end{eqnarray}
where $\Vcell$ is the volume of cell $\xi$, $\Vtot$ is the volume of
the root cell and $\alpha$, $\beta$ and $\eta$ are parameters
controlling the mesh refinement.

In the present work $\alpha$ and $\beta$ have been fixed to $1$, but
higher values can be useful for some particular situations where the
generated mesh must be tighter than the mesh generated directly from
the $\kappaabs$ and $T$ variations. Typically, these situations show
up for high optical depths (in this paper, optical depth values do not
exceed $\simeq 1$ at $10\,\microns$ along the rays). The practical
choice of $\alpha$, $\beta$ and $\eta$ is obtained from a compromise
between execution speed and numerical accuracy of the
Eq.~(\ref{eq:intensity}) integration (\eg typical values of $\eta$
range from $10^{-5}$ to $10^{-4}$). When dealing with optically thick
situations, a supplementary conditions can be added to
Eqs.~(\ref{eq:mesh1}) and~(\ref{eq:mesh2}) in order to prescribe an
upper limit to the cell optical depth. For instance, making use of the
computation of the integral in Eq.~(\ref{eq:mesh1}), one can add the
following criterion for cell $\xi$ (whose centre is
$(\rhocell,\zcell)$ and size $\Deltacell$)
\begin{equation}
  \label{eq:mesh3}
  \frac{1}{2\,\pi\,\rhocell\,\Deltacell}\,
  \iiint\limits_{\Vcell}\,\kappaext(\rr)\, d^3\rr \le \taulim \ ,
\end{equation}
 where $\taulim$ is the prescibed upper limit to the cell optical depth.

 For the moderate optical depths reached in this work, with values of
 $\eta$ down to $10^{-5}$ and $\taulim$ set to $10^{-2}$, the criteria
 of Eqs.~(\ref{eq:mesh1}) and~(\ref{eq:mesh2}) are the leading
 conditions to the mesh refinement.

\begin{figure}[!t]
  \centering
  \psfrag{x}{\Large $\pm \rho\,[\Rs]$}
  \psfrag{y}{\Large $z\,[\Rs]$}
  \psfrag{quadtree mesh}{\Large Quadtree mesh}
  \resizebox{\hsize}{!}{\includegraphics{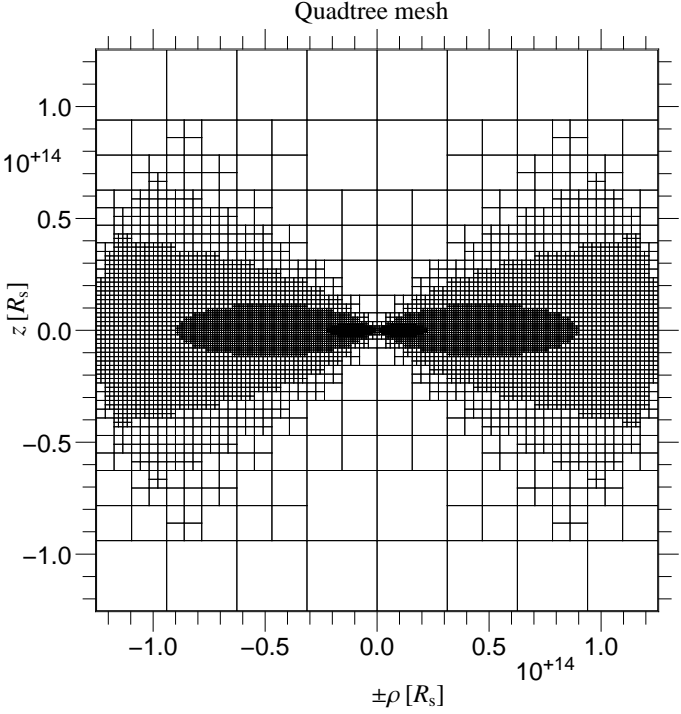}}
  \caption{Quadtree mesh for a disc configuration. The disc parameters
    are those of model~(b) described in Sect.~\ref{sect:models} (see
    also Tables~\ref{tab:param} and~\ref{tab:res}). The mesh
    refinement parameter $\eta$ (Eqs.~\ref{eq:mesh1}
    and~\ref{eq:mesh2}) has been set to the high value $10^{-3}$ in
    order to obtain a coarse mesh more easily represented.}
  \label{fig:mesh}
\end{figure}

 Figure ~\ref{fig:mesh} shows the mesh obtained in the particular case
 of a B[e] circumstellar disc (see Sect.~\ref{sect:testcase}) for
 models whose parameters are given in Table~\ref{tab:param} (see
 caption for more details).

The volume integrals in Eq.~(\ref{eq:mesh1}) and~(\ref{eq:mesh2}) are
estimated by Monte Carlo integration. For a quantity $f(\rr)$ and for
the cell $\xi$  the integral $\iiint\limits_{\Vcell}\,f(\rr) \,d^3\rr$
is approximated by
\begin{eqnarray}
  \label{eq:MC}
  2\pi\,
  \int\limits_{\zcell-\frac{\Deltacell}{2}}^{\zcell+\frac{\Deltacell}{2}}\,
  \int\limits_{\rhocell-\frac{\Deltacell}{2}}^{\rhocell+\frac{\Deltacell}{2}}\,
  \rho\,f(\rho,z)\,dz\,d\rho \approx
  \frac{2\,\pi\,\Deltacell^2}{N}\,
  \sum\limits_{k=1}^{N}\,\rho_k\,f(\rho_k,z_k) \ ,
\end{eqnarray}
where we made explicit use of the mesh coordinates and where
$(\rho_k,z_k)$ with $k=1,\cdots,N$ are chosen randomly and uniformly
within the cell domain.

\subsection{Symmetries}
\label{sect:symmetries}


We can make use of the CSE symmetries to reduce the
computation domain of an intensity map from Eq.~(\ref{eq:intensity})
to only a fourth of it and consequently reduce the computation time.

Recalling the definition of a ray (Eq.~\ref{eq:ray}), we have two
noticeable identities for any disc physical quantity $\Phi$ (\eg
$\kappaabs$, $\kappaext$, $T$, $n$, \ldots) depending on $\rr$
\begin{eqnarray}
  \label{eq:symmetries1}
  \Phi\left(\rr_s(X,Y,i)\right) &=&
  \Phi\left(\rr_{s}(-X,Y,i)\right)\quad\text{and} \\
  \label{eq:symmetries2}
   \Phi\left(\rr_s(X,Y,i)\right) &=&
   \Phi\left(\rr_{-s}(X,-Y,i)\right) \ ,
\end{eqnarray}
where Eq.~(\ref{eq:symmetries1}) expresses the disc symmetry with
respect to the $y-z$ plane and Eq.~(\ref{eq:symmetries2}) the point
symmetry with respect to the origin of the model system of
coordinates.

From the above identities it is straightforward to deduce their
counterpart for the intensity map
\begin{eqnarray}
  \label{eq:intsymm1}
  \Il(X,Y,i) &=& \Il(-X,Y,i) \quad\text{and} \\
  \label{eq:intsymm2}
  \Il(X,Y,i) &=& \Il(X,-Y,i)
  \e^{-\int\limits_{-\smax}^{+\smax}\,\kappaext(\rr_{s'})\,ds'} \ ,
\end{eqnarray}
where $\smax=\sqrt{\Rout^2-R^2}$. Note that the exponential factor in
Eq.~(\ref{eq:intsymm2}) has to be evaluated when computing $\Il(X,Y,i)$
anyway; no extra effort is required to derive $\Il(X,Y,i)$ from
$\Il(X,-Y,i)$ except for the multiplication of $\Il(X,-Y,i)$ by this
factor.

\subsection{Intensity map}

The fictitious image plane is split into a set of pixels whose
positions $X_j$ and $Y_k$ are given by
\begin{eqnarray}
  \label{eq:pixels}
  X_j &=& \Deltax\,\times\left(j+\frac{1}{2}-\frac{N}{2}\right)\ , \\
  Y_k &=& \Deltay\,\times\left(k+\frac{1}{2}-\frac{N}{2}\right) \ ,
\end{eqnarray}
where $\Deltax=\Deltay$ is the pixel size in $X$ and $Y$, and $N$ is
the number of pixels in $X$ and $Y$, and where
\begin{eqnarray}
  \label{eq:index}
  0 &\le& j,k \le (N\div 2)+\delta \ ,
\end{eqnarray}
where $\delta=-1$ for $N$ even and $\delta=0$ otherwise and ``$\div$''
stands for the integer division. Taking into account the symmetries
mentioned in Sect.~\ref{sect:symmetries} only a fourth of the pixels
need to be considered.

The evaluation of the integral in Eq.~(\ref{eq:intensity}) is carried
out for each pixel $(X_j,Y_k)$ and along the ray $\rr_s(X_j,Y_k,i)$.
The intersection points of the ray with the cell boundaries
corresponds to a set of distances along the ray defined as
\begin{eqnarray}
  \label{eq:distances}
  s_0 &=& 0 \\
  s_l &=& s_{l-1}+\Delta s_{l-1} \quad \quad \text{for $1 \le l \le \ncells$} \ ,
\end{eqnarray}
where $\ncells$ is the number of cells encountered along the ray, and
$\Delta s_l$ the distance crossed within the $l^\text{th}$ cell.

We estimate numerically the optical depth $\taul(X,Y,i;s)$, defined in
Eq.~(\ref{eq:tau}), via the midpoint rule quadrature by
\begin{eqnarray}
  \label{eq:taull}
  \taul(X,Y,i;s_l) \approx \taull =
  \sum\limits_{k=l}^{\ncells-1}\,\kappaext(\rr_{s_{l+1/2}})\,
  \Delta s_l \ ,
\end{eqnarray}
where we defined $s_{l+1/2}=s_l+\frac{\Delta s_l}{2}$ for
$l=0,\cdots,\ncells-1$.

The numerical estimate of $\Il(X_j,Y_k,i)$ is obtained by
\begin{eqnarray}
  \label{eq:Ijk}
  \Il(X_j,Y_k,i) \approx
  \sum\limits_{l=0}^{\ncells-1}\,
  \kappaabs(\rr_{s_{l+1/2}})\,\Bl(T(\rr_{s_{l+1/2}}))\,\e^{-\taull}\,\Delta s_l \ .
\end{eqnarray}

The results for all $j$ and $k$ can then be obtained from the discrete
counterpart of the symmetry relations~(\ref{eq:intsymm1})
and~(\ref{eq:intsymm2})
\begin{eqnarray}
  \label{eq:isymm1}
  \Il(X_{N-1-j},Y_k,i) &=& \Il(X_j,Y_k,i) \ , \\
  \label{eq:isymm2}
  \Il(X_j,Y_{N-k-1},i) &=& \Il(X_j,Y_k,i)\,\e^{-\taul^{(0)}} \ .
\end{eqnarray}

\subsection{Interferometric observables}

From the numerical estimate of $\Il(X_j,Y_k,i)$ given above we obtain
(similarly to Eqs.~\ref{eq:flux} and \ref{eq:visibility}) the
numerical fluxes and visibilities, which can be directly compared to
the observed data. The numerical estimate of these quantities is again
obtained through the mid-point rule.

The numerical flux $\Fl(i)$ is computed by
\begin{eqnarray}
  \label{eq:numflux}
  \Fl(i) \approx \frac{1}{d^2}\,
  \sum\limits_{k=0}^{N-1}\sum\limits_{l=0}^{N-1}\,\Il(X_k,Y_l,i)\,{\Deltax\Deltay} \ .
\end{eqnarray}
The complex visibility is approximated numerically by
\begin{eqnarray}
  \label{eq:numvis}
 \Vl \approx \frac{1}{d^2\,F_\lambda(i)}
  \sum\limits_{k=0}^{N-1}\sum\limits_{l=0}^{N-1}\,
  \Il(X_k,Y_l,i)\,\e^{
      2j\pi\frac{\vec{B}}{\lambda}\cdot\frac{\vec{R}_{kl}}{d}
  } \,{\Deltax\Deltay} \ ,
\end{eqnarray}
where $\vec{B}=(B\cos{\Delta},B\sin{\Delta})$ and
$\vec{R}_{kl}=(X_k,Y_l)$.

\subsection{Artificial data generation}
\label{sect:datagen}

The procedure described below aims to mimic the observables of the
VLTI/MIDI instrument: the flux $\Fl$ (Eq.~\ref{eq:numflux}) and the
modulus of the visibility $|\Vl|$ (Eq.~\ref{eq:numvis}). The
wavelengths and baselines chosen for the artificial data generation
correspond to accessible values to VLTI/MIDI with the Unit Telescopes
(UTs): $\lambda_j=7,8,9,10,11,12,$ and $13\,\microns$
($j=1,\cdots,\nlambda$; $\nlambda=7$), and $(\B_k,\PA_k)$ as shown in
Table~\ref{tab:BPA} ($k=1,\cdots,\nbase$; $\nbase=18$). These values
amount to $126$ points covering the uv-plane.

\begin{figure*}[!t]
  \centering
  $$
  \begin{array}{ccc}
    \psfrag{Mid-IR flux (model x)}{\Large Mid-IR flux (model a)}
    \psfrag{lambda}{\Large $\lambda\,[\microns]$}
    \psfrag{flux}{\Large $\Fl\,[\W\,\m^{-2}\,\microns^{-1}]$}
    \resizebox{80mm}{!}{\includegraphics{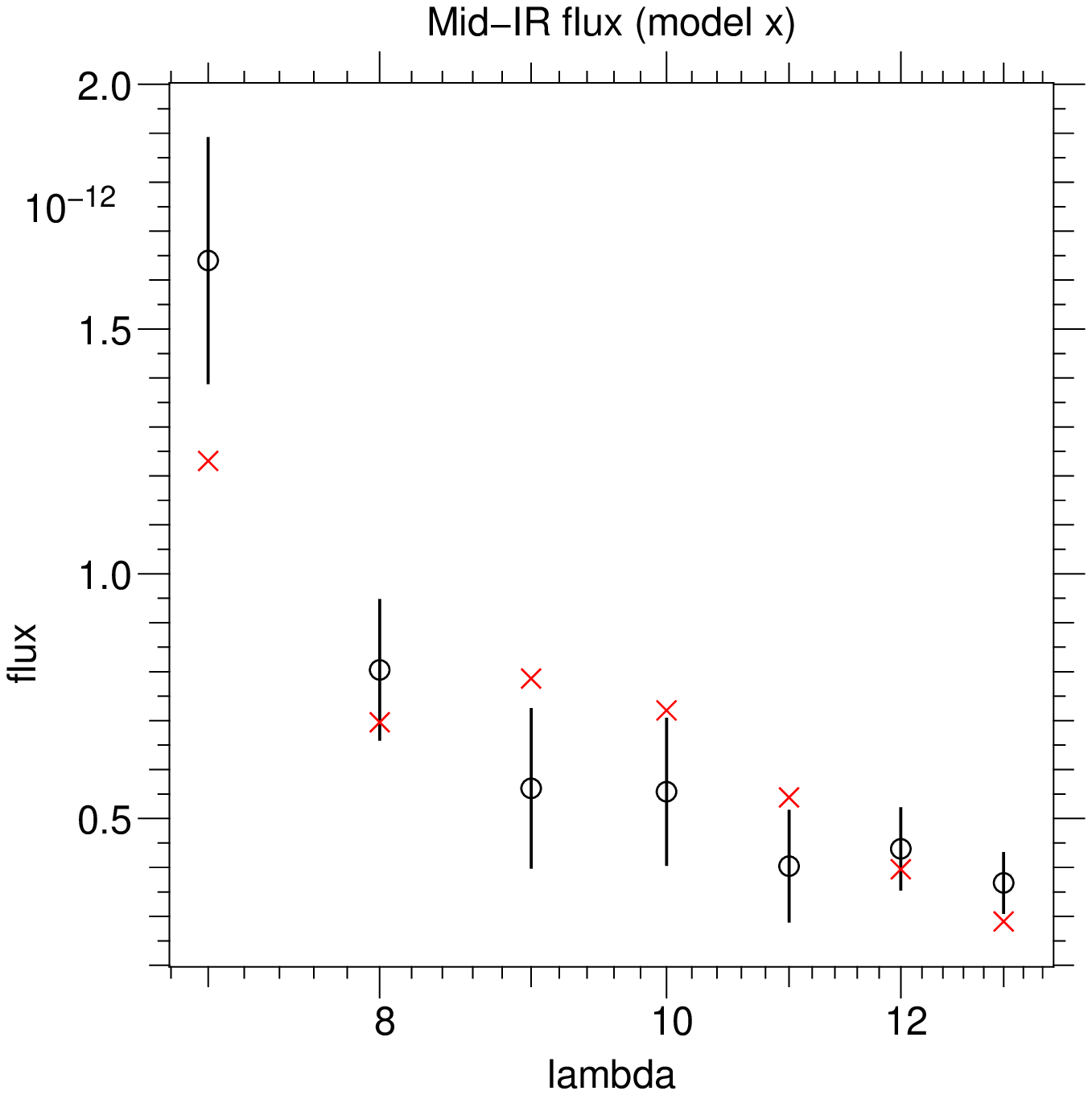}} & {\hspace{.5cm}} &
    \psfrag{Mid-IR flux (model x)}{\Large Mid-IR flux (model b)}
    \psfrag{lambda}{\Large $\lambda\,[\microns]$}
    \psfrag{flux}{\Large $\Fl\,[\W\,\m^{-2}\,\microns^{-1}]$}
    \resizebox{80mm}{!}{\includegraphics{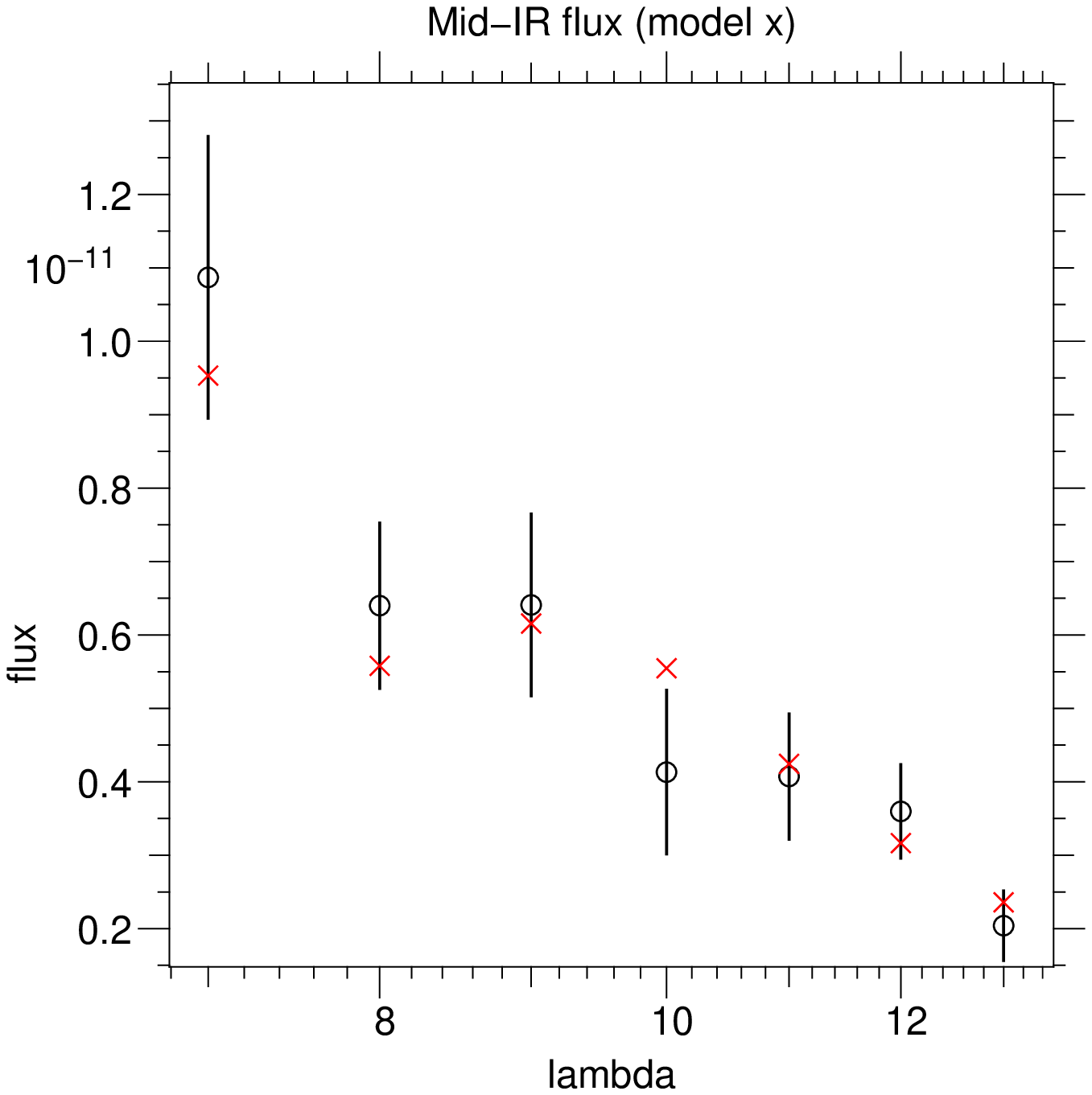}} 
  \end{array}
  $$
  \caption{Simulated VLTI/MIDI mid-IR object flux. The numerically
    generated data (left model~a, right model~b) are shown as circles
    with error bars. The values of the fluxes for the best-fit models
    are represented as crosses.}
  \label{fig:sed}
\end{figure*}

For a given intensity map at $\lambda_j$, $F_{\lambda_j}$ and
$|V_{\lambda_j}|$ are taken as the expectation values of the simulated
data. The observed flux $\Fobs_{\lambda_j}$ is then generated assuming
a Gaussian noise with an RMS (root mean square) corresponding to
$10\,\%$ relative error $\sigmaF(j)=0.1\times F_{\lambda_j}$.

The artificial observed visibility amplitudes $|\Vobs_\lambda|$ are
obtained as
\begin{eqnarray}
  \label{eq:Vobs}
  |\Vobs_{\lambda_j}(\B_k,\PA_k)|=|V_{\lambda_j}(\B_k,\PA_k)|+\Delta V_k \ ,
\end{eqnarray}
where $\Delta V_k$ is a wavelength independent shift that mimics the
error in the observed visibilities, introduced by the calibration
procedure commonly used in optical/IR interferometry. For each
$(\B_k,\PA_k)$, $\Delta V_k$ is computed assuming a Gaussian noise
with an RMS corresponding to $10\,\%$ relative error (typical for
VLTI/MIDI) $\sigmaV(k)=0.1\times\langle|V_\lambda(\B_k,\PA_k)|
\rangle$, where $\langle|V_\lambda(\B_k,\PA_k)| \rangle$ is the
wavelength mean visibility modulus.

\begin{table}[!t]
  \caption{Projected baselines. These values correspond to the baselines accessible from pairs of Unit Telescopes (UT) at ESO-VLTI.}
  \label{tab:BPA}
  \centering
  \begin{tabular}{ccc}
    \hline\hline
    $k$ & $B_k\,[\m]$ & $\PA_k\,[\mathrm{deg}]$ \\
    \hline
    1  & 37.8  & 61.7 \\
    2  & 41.3  & 53.4 \\
    3  & 43.7  & 44.8 \\
    4  & 46.2  & 44.5 \\
    5  & 49.5  & 37.5 \\
    6  & 51.9  & 30.0 \\
    7  & 61.7  & 134.6 \\
    8  & 62.0  & 111.2 \\
    9  & 62.4  & 122.5 \\
    10 & 81.3  & 108.2 \\
    11 & 83.0  & 52.2 \\
    12 & 86.3  & 96.0 \\
    13 & 89.0  & 84.4 \\
    14 & 89.9  & 44.8 \\
    15 & 94.8  & 36.7 \\
    16 & 113.6 & 82.4 \\
    17 & 121.2 & 73.6 \\
    18 & 126.4 & 64.9  \\
    \hline
  \end{tabular}
\end{table}


\subsection{Model fitting and error estimate}
\label{sect:constraints}

We describe here the procedure adopted in order to simultaneously fit
observed fluxes and visibilities using FRACS models defined by a given
set of input parameters. This procedure is applied to artificial data
in the next sections.

In order to quantify the discrepancy between the artificial
observations ($|\Vobs_{\lambda_j}|$ and $\Fobs_{\lambda_j}$) and the
visibilities and fluxes from a given model
($|V_{\lambda_j}(\B_k,\PA_k)|$ and $F_{\lambda_j}$) we use the
$\chi^2$ like quantities
\begin{eqnarray}
  \label{eq:chisV}
  \chisV =\sum\limits_{j=1}^{\nlambda}\!\sum\limits_{k=1}^{\nbase}\,
  \left(\frac{|\Vobs_{\lambda_j}(\B_k,\PA_k)|-|V_{\lambda_j}(\B_k,\PA_k)|}{\sigmaV(k) }\right)^2 \ ,
\end{eqnarray}
and
\begin{eqnarray}
  \label{eq:chisF}
  \chisF=\sum\limits_{j=1}^{\nlambda}\,\!\sum\limits_{k=1}^{\nbase}\,
  \left(\frac{\Fobs_{\lambda_j}-F_{\lambda_j}}{\sigmaF(j)}\right)^2 \ .
\end{eqnarray}


To take into account both the mid-IR flux and the visibilities on the
same level in the fitting process, we minimise the following sum
\begin{eqnarray}
  \label{eq:chi2}
  \chi^2 = \chisV + \chisF \ .
\end{eqnarray}


In the discussion below about the parameter and error determination we
use the reduced $\chi^2$ defined by
$\chir^2=\chi^2/(2\nbase\nlambda-n_{\rm free})$ (for $n_{\rm free}$
free parameters).



From a minimising algorithm the best-fit model parameters can then be
found by determining the minimum $\chir^2$: $\chirmin^2$. The
``error'' estimate is obtained from a thorough exploration of the
parameter space volume, defined by a contour level $\chirmin^2
+\Delta\chir^2$, where $\Delta \chir^2$ has been chosen equal to 1.
This volume can be interpreted as a confidence region. The quantity
defined in Eq.~(\ref{eq:chi2}) is a weighted sum of $\chi^2$ variables
whose cumulative distribution function can be approximated by a gamma
distribution~\citep[see][]{feiveson1968} with the same mean and
variance. It is then possible to obtain a rough estimate of the
confidence level associated with the $\Delta \chir^2=1$ confidence
region given approximately by $\simeq 2\,\sigma$.








The size of the confidence region is determined by considering all
possible pairs of parameters for a given fitted model and computing
$\chir^2$ maps for each. The procedure to estimate the errors can be
summarised as follows:
\begin{itemize}
\item For a given $\chir^2$ map, \ie for a given couple of parameters
  among the $n_{\rm free}\times(n_{\rm free}-1)/2$ possibilities, we
  identify the region bounded by $\Delta \chir^2=1$ around the minimum
  of this particular map.
\item The boundaries of the projection of these regions on each of the
  two parameter axis considered are recorded for each map.
\item The final errors on a given parameter are taken as the highest
  boundary values of the projected regions over all maps.
\end{itemize}

\section{Astronomical test case: sgB[e] stars}
\label{sect:testcase}


In the following sections we apply FRACS to a theoretical
interferometric study of dusty CSE of B[e] supergiants (sgB[e] in the
nomenclature of~\citealt{lamers1998}). However, we emphasise that
FRACS is in no way restricted to this particular class of objects.

sgB[e] stars reveal in particular a strong near- or mid-IR excess
caused by hot dust emission. There is
evidence~\citep[e.g.][]{zickgraf1985} that the stellar environment,
and in particular dust, could be confined within a circumstellar
disc. Our purpose is to characterise this class of objects and derive
not only geometrical parameters (\eg inner dust radius, disc position
angle and inclination) but also physical parameters such as
temperature gradients, dust formation region, material density, \ldots






The physical description of the CSE chosen for our study is the wind
model with equatorial density enhancement. This is a classical CSE
model commonly adopted for sgB[e] \citep[e.g.][]{porter2003}.

In order to compute the model intensity maps we need to parameterise
the emissivity of the disc. Consistently with FRACS assumptions, we
consider only dust thermal emission without scattering by dust grains
and the gas contribution to the medium emissivity. In the rest of this
section we characterise the emissivity by describing the dust density
law, the absorption cross section, and the temperature structure of
the CSE.

\subsection{Mass loss and dust density}

Dust is confined between the inner and outer radius $\Rin$ and $\Rout$
respectively. We assume a stationary and radial mass loss; physical
quantities will consequently depend only on the radial component $r$
and the co-latitude $\theta$. The disc symmetry axis coincides with
the $z$ axis of the model cartesian system of coordinates. The mass
loss rate and velocity parametrisations are simplifications of the one
adopted by~\citet{carciofi2010}, and we refer the reader to their work
for a complete description~\citep[see also][for a similar
description]{stee1995}.

The mass loss rate per unit solid angle, at co-latitude $\theta$, is
parameterised as follows
\begin{equation}
  \label{eq:massLossRate}
  \dMdO(\theta) = \dMdO(0)\left(1+A_1\,\sin^m(\theta)\right) \ ,
\end{equation}
with the help of two dimensionless parameters $A_1$ and $m$.


Even though our computations make no explicit use of the radial
velocity field $v_r(\theta)$ (assumed to have reached the terminal
velocity $v_\infty(\theta)$ in the region under considerations, \ie
$v_r(\theta) \approx v_\infty(\theta)$), the dust density depends on
$v_r(\theta)$ parameterised in a similar fashion
\begin{equation}
  \label{eq:velocity}
  v_r(\theta) = v_r(0)\,\left(1+A_2\,\sin^m \theta \right) \ ,
\end{equation}
where we have introduced the supplementary dimensionless parameters
$A_2$. From Eqs.~(\ref{eq:massLossRate}) and~(\ref{eq:velocity}) we
see that $A_1$ and $A_2$ are the relative differences of the values of
$\dMdO(\theta)$ and $v_r(\theta)$ at the equator and the pole
(relatively to the pole).

From the mass continuity equation one obtains the number density of
dust grains
\begin{equation}
  \label{eq:density}
  n(r,\theta)=\rhoin\,\left(\frac{\Rin}{r}\right)^2\,
  \frac{1+A_2}{1+A_1}\,
  \frac{1+A_1\,\left(\sin{\theta}\right)^m}{1+A_2\,\left(\sin{\theta}\right)^m} \ ,
\end{equation}
where $\rhoin$ is the dust grain number density at $\Rin$ in the disc
equatorial plane. In Eq.~\ref{eq:density}, the parameter $m$ controls
how fast the density drops from the equator to the pole, defining an
equatorial density enhancement (disc-like structure).


Consistent with the accepted conditions for dust formation
\citep{carciofi2010,porter2003} we assume that the dust can survive
only in the denser parts of the disc. We thus define a dusty disc
opening angle $\dthetadisc$ determined by the latitudes for which the
mass loss rate has dropped to half of its equatorial value:
\begin{equation}
  \label{eq:thetadust}
  \dthetadisc=
  2\,\arccos{\left(\frac{A_1-1}{2\,A_1}\right)^{\frac{1}{m}}} \ .
\end{equation}

To summarise, the dust grains only exist (i.e., $n(r,\theta)\neq0$) in
the regions bounded by $\Rin \leq r \leq \Rout$ and by
$\frac{\pi-\dthetadisc}{2} \leq \theta \leq
\frac{\pi+\dthetadisc}{2}$.

\subsection{Dust opacities}

The absorption cross section $\Cabs$ for the dust grains is obtained
from the~\citet{mie1908} theory. The Mie absorption cross sections are
computed from the optical indices of astronomical
silicate~\citep{draine1984}. Note that since scattering is neglected,
$\Cabs\approx\Cext$, with $\Cext$ being the extinction cross section.

For a power-law size distribution function according
to~\citet{mathis1977} the mean cross sections (\eg for $\Cabs$) are
given by
\begin{eqnarray}
  \label{eq:crosssection}
  \Cabs = \frac{\int\limits_{\amin}^{\amax}\,a^{-\beta}\,\Cabs(a)\,da}
  {\int\limits_{\amin}^{\amax}\,a^{-\beta}\,da} \ ,
\end{eqnarray}
where $\amin$ and $\amax$ are the minimum and maximum radii for the
dust grains under consideration and $\beta$ is the exponent of the
power-law. The computation of the cross section in
Eq.~(\ref{eq:crosssection}) was performed with the help of
the~\citet{wiscombe1980} algorithm.

\subsection{Temperature structure}
\label{sect:tempstruct}


The dust temperature is assumed to be unique (\ie independent of grain
size) and described by a power-law
\begin{eqnarray}
  \label{eq:temperature}
T(r)=\Tin\,\left(\frac{\Rin}{r}\right)^{\gamma}\ ,
\end{eqnarray}
where $\Tin$ is the temperature at the disc inner radius $\Rin$. We
note that $\gamma$ is not necessarily a free parameter because in the
optically thin regime (large wavelength and radius) the temperature
goes as $T(r)\propto r^{-\frac{2}{4+\delta}}$ with $\delta \simeq
1$~\citep[see][]{lamers1999}.


\subsection{Central continuum emission}

The continuum emission from the central regions is composed by the
emission from the star and from the close ionised gas (free-free and
free-bound emission). This central source emission is confined to a
small region of radius $\Rs$ ($\ll\Rin$), which is unresolved (angular
sizes of a few milliarcseconds) by mid-IR interferometers. Thus, in
our modelling $\Rs$ is simply a scaling factor of the problem fixed to
a typical radius value for massive stars. The specific intensity (in
$\W\,\m^{-2}\,\microns^{-1}\,\str^{-1}$) of this central source is
parameterised as follows
\begin{eqnarray}
  \label{eq:csource}
  \Isl = \phis\,\left(\frac{\lambda_0}{\lambda}\right)^\alpha \ ,
\end{eqnarray}
where $\phis$ is the specific intensity at a reference wavelength
$\lambda_0$ ($=10\,\microns$ in the following), and $\alpha$ gives the
spectral dependence of the continuum radiation. In the mid-IR its
value is expected to lie between $\alpha=4$ (pure black body) and
$\alpha\simeq 2.6$ (free-free emission) for an electron density
proportional to $r^{-2}$ \citep{panagia1975,felli1981}.

\section{Study of the tested models}
\label{sect:numtests}

\begin{figure*}[f]
  \centering
  \includegraphics[width=17cm]{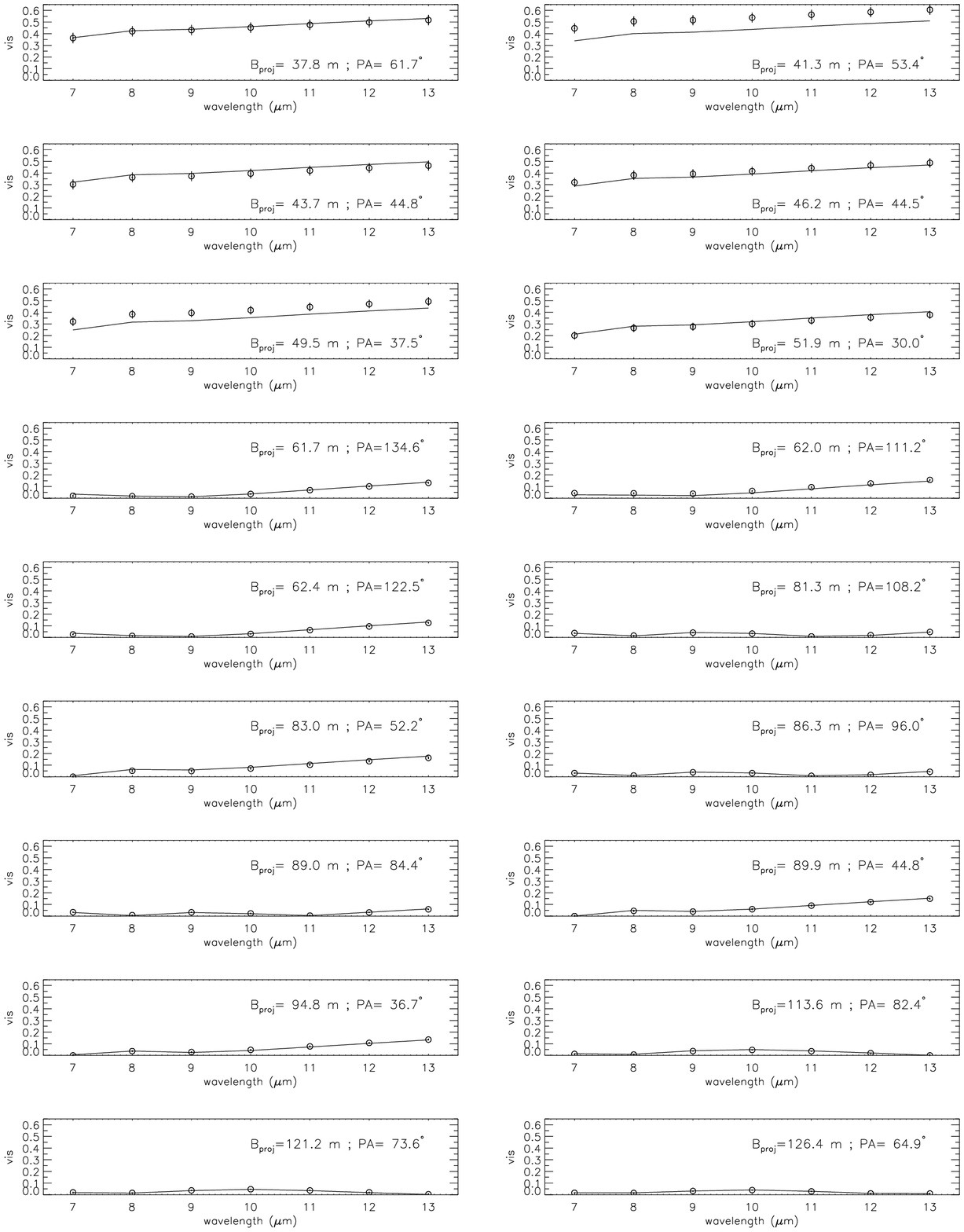}
  \caption{Visibilities of the artificial sgB[e] circumstellar
    environment (model~a). The visibility variations with the
    wavelength are shown for each baseline specified by the value of
    the projected baseline and the position angle on the sky. The
    circles represent the simulated observations, and the solid curves
    represent the best-fit model.}
  \label{fig:visib1}
\end{figure*}

\begin{figure*}[f]
  \centering
  \includegraphics[width=17cm]{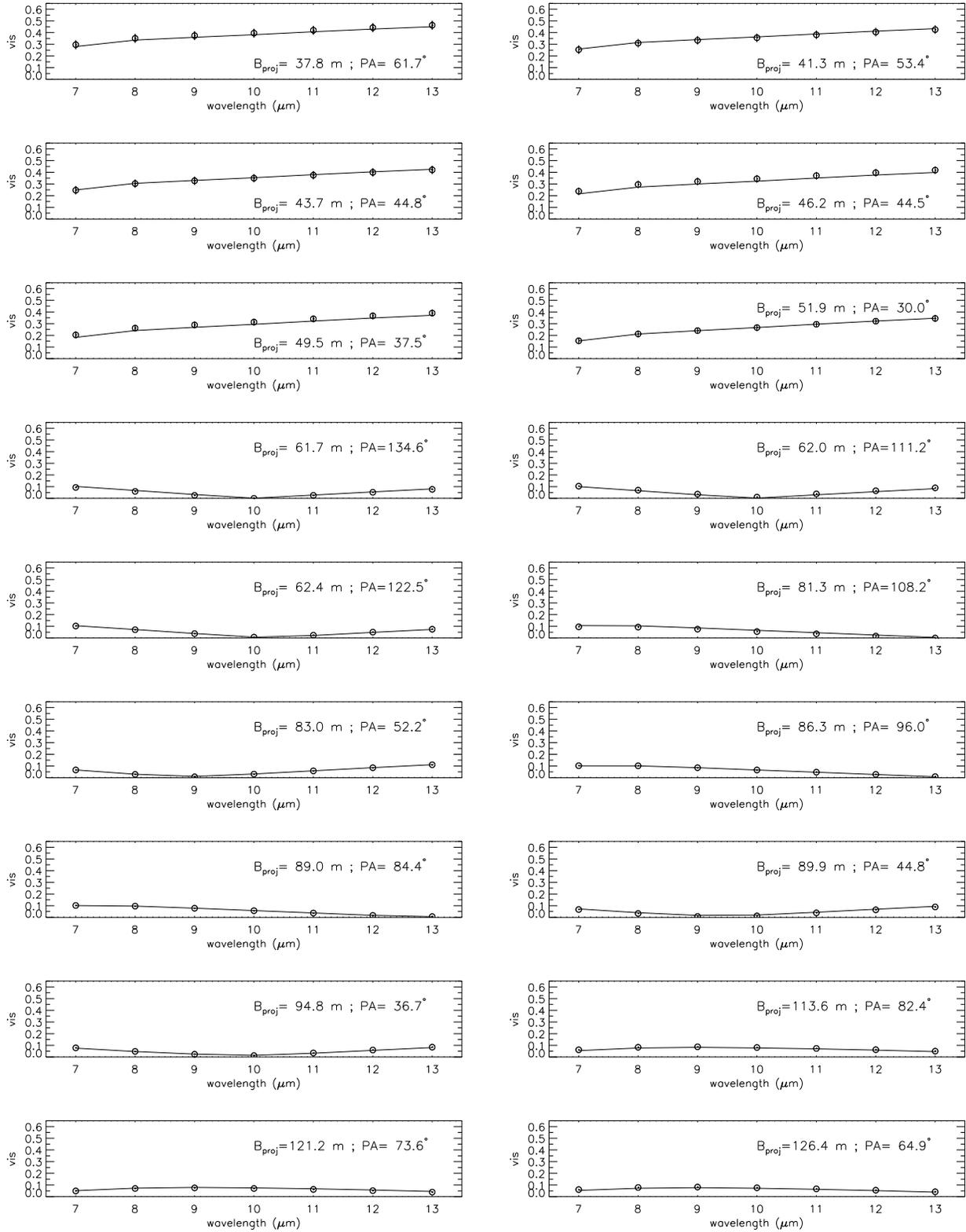}
  \caption{Visibilities of the artificial sgB[e] circumstellar
    environment (model~b).}
  \label{fig:visib2}
\end{figure*}



\begin{table}[!t]
  \caption{Model parameters. This table lists the parameters of 12 different models. The parameter values that change ($\dthetadisc$, $\rhoin$ and $i$) from one model to the other have been inclosed in a box and separated by a slash. The values of $\dthetadisc=10\,\degr/60\,\degr$ given below correspond via Eq.~(\ref{eq:thetadust}) to $m=183.56/4.86$ respectively.}
  \label{tab:param}
  \centering
  \begin{tabular}{ccc}
    \hline\hline
    Parameters & Values & Unit \\
    \hline
    $A_1$ & $150$ & - \\
    $A_2$ & $-0.8$ & - \\
    $\dthetadisc$ & \fbox{$10/60$} & deg \\
    $\Rs$ & $60$ & $\Rsun$ \\
    $\Rin$ & $30$ & $\Rs$ \\
    $\Rout$ & $3000$ & $\Rs$ \\
    $\rhoin$ & \fbox{$0.015/0.15$} & $\m^{-3}$ \\
    $\Tin$ & $1500$ & $\K$ \\
    $\gamma$ & $0.75$ & - \\
    $\phis$ & $6500$ & $\W\,\m^{-2}\,\microns^{-1}\,\str^{-1}$ \\
    $\alpha$ & $3$ & - \\
    $\PAd$ & $125$ & deg \\
    $i$ & \fbox{$20/50/90$} & deg \\
    $\amin$ & $0.5$ & $\microns$ \\
    $\amax$ & $50$ & $\microns$ \\
    $\beta$ & $-3.5$ & - \\
    \hline
  \end{tabular}
\end{table}

Following the description in the last sections we describe here the
chosen sgB[e] model parameters used to simulate VLTI/MIDI observations
(visibilities and fluxes) and the corresponding analysis, \ie model
fitting, using FRACS. The list of chosen parameters is summarised in
Table~\ref{tab:param}. Two types of numerical tests are
presented. Firstly, synthetic mid-IR interferometric data are
generated from FRACS itself. In that way, it is possible to determine
what information the mid-IR interferometric data contain under the
optimistic assumption that we do have the {\em true} model. Secondly,
this study is supplemented by the comparison of FRACS to a Monte Carlo
radiative transfer computation. This confirms that FRACS can indeed
mimic, under appropriate conditions, the results of a more
sophisticated code as seen from the mid-IR interferometric eye.

\subsection{Parameter description}

The distance to the simulated object has been fixed to $d=1\,\k\pc$,
which is a typical distance for Galactic sgB[e].


The inner radius $\Rin=30\,\Rs=1800\,\Rsun$ value was chosen by
considering the location of the hottest dust
grains~\citep[see][]{lamers1999} with a condensation temperature of
$1500\,\K$ assumed to be the $\Tin$ value. The value of $\Rout$ cannot
be determined from the mid-IR data and has been fixed to
$3000\,\Rs=1.8\times 10^5\,\Rsun$. The temperature gradient $\gamma$
was fixed to $0.75$ according to~\citet{porter2003}.  $\PAd$ was fixed
to $125\,\degr$.

The central source emission is supposed to have a radius
$\Rs=60\,\Rsun$. We recall that the central region is unresolved by
the interferometer and that its radiation describes both the stellar
and inner gas contribution to the continuum mid-IR emission. The
specific intensity of this central source $\phis$ has been chosen to
be $6500\,\W\,\m^{-2}\,\microns^{-1}\,\str^{-1}$. If the central
source was a pure blackbody this value would correspond to the
$10\,\microns$ emission of a blackbody with an effective temperature
around $\simeq 8000\,\K$. However, this central emission is not a pure
blackbody, and we adopt the spectral dependence of the central source
emission to be $\alpha=3$, which is a compromise between $\alpha=4$
for a pure blackbody and a value of $\simeq 2.6$ for free-free
emission~\citep{panagia1975,felli1981}.

Spectroscopic observations of $\H\alpha$ and forbidden line emissions
from B[e] CSE~\citep{zickgraf2003} reveal that typical values for
$A_2$ are expected to range from $-0.95$ to $-0.75$. We adopt the
value $-0.8$ in our models. According to~\citet{lamers1987}, the
values of $A_1$ range from $10^2$ to $10^4$ in most cases (though
values as low as $10$ are not excluded). With this high value of $A_1$
the factor $1+A_1(\sin{\theta})^{1/m}$ in Eq.~(\ref{eq:density}) of
$n(r,\theta)$ is approximatively given by $A_1(\sin{\theta})^{1/m}$
for all pertinent values of $\theta$, \ie those close to $\pi/2$
within the disc. This leads to an evident degeneracy in $\rhoin\times
A_1$ in $n(r,\theta)$: we are only sensible to the product of the two
parameters as a scaling factor for the density. Therefore, the value
of $A_1$ is assumed to be fixed to 150.


To define the dust opacities the chosen value for $\beta$ is that
of~\citet{mathis1977}, \ie $\beta=-3.5$. Because some sgB[e] show weak
$9.7\,\microns$ silicate features in their spectrum
\citep[e.g.][]{porter2003,domiciano2007} we chose to use large grains
in our test models: $\amin=0.5\,\microns$ and
$\amax=50\,\microns$. However, with this particular choice of large
grains, the average albedo from $7$ to $13\,\microns$ is $6.4\,\%$,
with the highest value reached at $7\,\microns$. We have checked with
a Monte Carlo (MC) simulations (see Sect.~\ref{sect:tapas}) that the
effect of scattering on our primary observables, visibilities, and
fluxes is indeed negligeable by comparing the results obtained by
switching the scattering process off and on \footnote{The computation
  have been done for model~b described in Sect.~\ref{sect:models}, the
  baselines listed in Table~\ref{tab:BPA} and the wavelengths under
  consideration from $7$ to $13\,\microns$.}. The mean relative
differences are $3.5\,\%$ and $3.0\,\%$ for the visibilities and the
fluxes respectively. These values must be compared to the effect of
random noise in the MC simulation, estimated to be of the same order
and to experimental errors, typically $\sim 10\,\%$ for the
visibilities and fluxes. We underline that whenever the albedo can be
neglected, it is theoretically safe to compute visibilities and fluxes
from the consideration presented in Sect.~\ref{sect:raytracing}, in
any other situations the effect of scattering on the observable must
be carefully tested.

The parameters $\rhoin$, $m$ and $i$ were set to different values
defining 12 test models to be analysed from their corresponding
simulated data. Two $\rhoin$ values ($0.015$~$\m^{-3}$ and
$0.15$~$\m^{-3}$) have been chosen in order to have an approximate
disc-dust optical depth in the equatorial plane (from $\Rin$ to
$\Rout$) close to $\simeq 0.1$ and $\simeq 1$ in the wavelength range
considered (from $7\,\microns$ to $13\,\microns$).  These values
corresponds to a mass loss rate of $\dot{M} \simeq 2.5\times 10^{-7
  \cdots -6}\,\msunyr$. Two $m$ values were chosen corresponding to a
wide and a narrow opening angle, \ie $\dthetadisc=10\,\degr$ and
$\dthetadisc=60\,\degr$. Three inclinations $i$ were tested
($20\degr$, $50\degr$, and $90\degr$) corresponding to discs seen
close to pole-on, intermediate inclination, and equator-on. These
values of $\rhoin$, $m$, $i$, together with the parameters fixed
above, define 12 test models that will be studied below.

From these 12 test models we have generated 12 sets of artificial
VLTI/MIDI observations (visibilities and fluxes) following the
procedure described in Sect.~\ref{sect:datagen}. We do not aim to
present an exhaustive revue of all types of sgB[e] CSE. Rather, we
focussed on the analysis of the parameter constraints one can hope to
obtain from present and near-future mid-IR spectro-interferometry. The
quantitative estimate of these constraints is derived from a
systematic analysis of the $\chir^2$ variations with the parameters.

In our model fitting and $\chir^2$ analysis we concentrate on 10~free
parameters ($n_{\rm free} = 10$) that can be set into four different
groups:
\begin{itemize}
\item The {\sl geometrical} parameters~: $\PAd$, $i$ and $\Rin$,
\item the parameters related to the {\sl central source}~: $\phis$ and
  $\alpha$,
\item those describing the {\sl temperature structure}~: $\Tin$ and
  $\gamma$,
\item and the {\sl number density of dust grains}~: $A_2$,
  $\dthetadisc$ (or equivalently $m$) and $\rhoin$.
\end{itemize}

The remaining parameters of the model are in general loosely
constrained by mid-IR interferometric observations so that we kept
them fixed to the values described above.

\begin{table*}[!t]
  \caption{Relative errors (given in $\%$) on the parameters for models~(a) et (b) (see text for description of models). For each of the 10 free parameters considered in the analysis, the values of the relative error corresponding to the 12 different models are given. Indeed, these relative errors are ``mean values'' for the errors because the error bars are not symmetric with respect to the best-fit values. The parenthesis around the relative error of $A_2$ recall that this parameter is bounded. }
  \label{tab:errors}
  \centering
  \begin{tabular}{cllllllllll}
    \hline\hline
    Models$\backslash$Parameters & $A_2$ & $m$  & $\Rin$  & $\rhoin$  & $\Tin$  & $\gamma$  & $\phis$  & $\alpha$  & $\PAd$  & $i$ \\ \hline
    {\bf (a)} & (53) & $\ge 100$ & 1.9 & 46  & 15 & 7.1 & 27        & 13 & 4.3 & 6.2 \\
    {\bf (b)} & (59) & $\ge 100$ & 4.5 & 100 & 20 & 12  & $\ge 100$ & 96 & 9.4 & 7.2 \\
    \hline
  \end{tabular}
\end{table*}

 \begin{table*}[!t]
   \caption{Constraints on the model parameters. For the 12 models considered here (differing in their value of $\tau$, $\dthetadisc$ and $i$), numbered from 1 to 10 (except for model a and b), we classified the parameters into 3 different relative error ranges~: below $10\,\%$, between $10$ and $25\,\%$ and above $25\,\%$. Because $A_2$, $m$, $\rhoin$, $\phis$ are determined for all the models with an error greater than 25 \% they have been discarded from the table for the  sake of clarity .}
  \label{tab:res}
  \centering
  \begin{tabular}{clcc|lll}
    \hline\hline
    \multicolumn{4}{c|}{Models$\backslash$parameters} & \multicolumn{3}{|c}{Constraints [\%]} \\
    & $\tau$ & $\dthetadisc\,[\text{deg}]$ & $i\,[\text{deg}]$
    & $\le 10$ & $10\rightarrow 25$ & $\ge 25$ \\ \hline
    1   & 0.1 & 60 & 20 & $\Rin$, $\gamma$ & $\Tin$, $\alpha$ &   $\PAd$, $i$ \\
    2   & 1.  & 60 & 20 & $\Rin$ & $\Tin$, $\gamma$ &   $\alpha$, $\PAd$, $i$ \\
    3   & 0.1 & 10 & 20 & - & $\Rin$ &  $\Tin$, $\gamma$, $\alpha$, $\PAd$, $i$ \\
    4   & 1.  & 10 & 20 & $\Rin$, $\gamma$ & $\Tin$  &   $\alpha$, $\PAd$, $i$ \\
    (a) & 0.1 & 60 & 50 & $\Rin$, $\gamma$, $\PAd$, $i$ &  $\Tin$, $\alpha$&    \\
    (b) & 1.  & 60 & 50 & $\Rin$, $\PAd$, $i$ & $\Tin$, $\gamma$ &   $\alpha$  \\
    5   & 0.1 & 10 & 50 & - & $\Rin$ & $\Tin$, $\gamma$, $\alpha$, $\PAd$, $i$   \\
    6   & 1.  & 10 & 50 & $\Rin$, $\gamma$, $\PAd$, $i$ & $\Tin$&   $\alpha$ \\
    7   & 0.1 & 60 & 90 & $\Rin$, $\gamma$, $\PAd$ & $\Tin$, $i$, $\alpha$  \\
    8   & 1.  & 60 & 90 & $\Rin$, $\PAd$ &  $\Tin$, $\gamma$, $i$ &   $\alpha$ \\
    9   & 0.1 & 10 & 90 & $\PAd$ & $\Rin$ & $\Tin$, $\gamma$, $\alpha$, $i$  \\
    10  & 1.  & 10 & 90 & $\Rin$, $\gamma$, $\PAd$, $i$ &  $\Tin$ &   $\alpha$ \\ \hline
   \end{tabular}
\end{table*}

\subsection{Model fitting and $\chir^2$ analysis of the 12 test models}
\label{sect:models}

We describe here the data analysis procedure adopted to study our 12
test models. The results of our analysis are summarised in
Tables~\ref{tab:errors} and~\ref{tab:res}, and their physical
interpretation is presented in Sect.~\ref{sect:discussion}.

As a first step we chose 2 of the 12~models, hereafter called models
(a) and (b), to be exhaustively studied from a complete model fitting
procedure. As an example we show the simulated observed mid-IR fluxes
and visibilities for model~(a) and~(b) in Figs.~\ref{fig:sed},
\ref{fig:visib1}, and~\ref{fig:visib2}. The parameters of models (a)
and (b) are those of Table~\ref{tab:param} with
$\dthetadisc=60\,\degr$, $i=50\,\degr$ and $\rhoin$ fixed to the
values $0.015\,\m^{-3}$ and $0.15\,\m^{-3}$ respectively. These two
models are those presenting some of the best constrained model
parameters for the dust CSE. On the other hand, the contribution of
the central regions to the total flux and visibilities is quite
different in models (a) and (b) (see discussion in
Sect.~\ref{sect:discussion}).


The study of models (a) and (b) have thus been performed as for real
interferometric observations. The best-fit values of the parameters
have been obtained by the Levenberg-Marquardt algorithm with a
stopping criterion corresponding to a relative decrease in $\chir^2$
of $10^{-3}$.

The errors on each model parameter have been obtained following the
methods described in Sect.~\ref{sect:constraints}. The $\chir^2$ maps
have been computed with a resolution of $21\times 21$ around the
best-fit values of the parameters. The map sizes have been adjusted in
order to enclose the $\Delta \chir^2=1$ contour. This adjustment was
performed until an upper limit for the map size of $100\,\%$ of the
best-fit parameter values was reached. This amounts to the computation
of $3.969\times 10^{4}$ different models. The results, namely the mean
relative error up to $100\,\%$, for these two particular models are
summarised in Table~\ref{tab:errors}.

The other ten models (numbered from~1 to~10 in Table~\ref{tab:res})
have been used in order to get some quantitative (but limited)
information about how the uncertainties of the fitted parameters
evolve as a function of three disc characteristics: its optical depth
($\tau$ by means of $n_{in}$ parameter), its inclination ($i$) and its
opening angle ($\dthetadisc$, controlled by $m$). To perform this
study we have decided to limit the exploration of the space parameter
in a relative range of 25 \% on both sides from the model
parameters. In order to reduce the computation time, the maps were not
generated around the best-fit parameters which would have required to
compute several thousands models more but around the true parameters
themselves. This procedure has the supplementary advantage that we do
not rely on any specific minimisation algorithm. We checked that
estimating the best-fit parameters from the true ones is reasonable
within a few percents using the Levenberg-Marquardt algorithm with a
stopping criterion corresponding to a relative decrease in $\chir^2$
of $10^{-3}$. The resolution of the $\chir^2$ maps have been reduced
to $15\times15$.  The total number of models to be computed is as
large as $1.0125\times 10^5$.




\subsection{Comparison with a Monte Carlo simulation}
\label{sect:tapas}

\begin{table*}[t]
  \begin{minipage}[t]{\textwidth}
    \caption{best-fitting FRACS parameters from artificial data
      generated with a Monte Carlo code. The column ``true values''
      refers to the MC input parameters, except for $\Tin$, $\gamma$,
      and $\gamma'$ which are determined from the results of the MC
      simulation. The columns ``two power-law'' and ``one power-law''
      list the best-fit parameters obtained with FRACS assuming two
      and one power-law for the temperature respectively. The
      $\chirmin^2$ values are respectively $0.73$ and $0.79$ for two
      and one power-law.}
    \label{tab:fracsvstapas}
  \centering
  \renewcommand{\footnoterule}{}
  \renewcommand{\thefootnote}{\thempfootnote}
  \begin{tabular}{ccccc}
    \hline\hline
    Parameters & Units & true values & two power-law & one power-law \\
    \hline
    $A_2$ & - & -0.8 &-0.791 & -0.782 \\
    $m$ & - & 4.86 & 5.59 & 4.74 \\
    $\Rin$ & $\Rs$ & 30 & 29.8  & 29.9 \\
    $\rhoin$ &  $\m^{-3}$ & 0.15 & 0.189 & 0.169 \\
    $\Tin$ & $\K$ & 1150\footnote{These values are not prescribed parameters, but are determined from the results of the Monte Carlo simulation. The values reported here are best-fit parameters of the mean disc temperature (see text for more details).} & 1090  & 1070 \\
    $\gamma/\gamma'$ & - & 0.725/0.478\footnotemark[\value{mpfootnote}] & 0.719/0.613 & 0.676 \\
    $\RT$ & $\Rin$ & 5.24\footnotemark[\value{mpfootnote}] & 2.87 & - \\ 
    $\phis$ & $\W\,\m^{-2}\,\microns^{-1}\,\str^{-1}$ & $6.62\times 10^3$ & $6.48\times 10^3$  & $5.04\times 10^3$ \\
    $\PAd$ & deg & 125  & 125 & 124 \\
    $i$ & deg & 50.0 & 50.6  & 50.2 \\
    \hline
  \end{tabular}
\end{minipage}
\end{table*}

We generated synthetic data with the help of a Monte Carlo (MC)
radiative transfer code~\citep{niccolini2006} for model~b (see above)
for the seven wavelengths considered in the problem and the baselines
of Table~\ref{tab:BPA}. Again, the adopted procedure to generate the
mid-IR interferometric data follows the considerations of
Sect.~\ref{sect:datagen}. In the MC code, the source of photons is
described by a blackbody sphere of radius $\Rs=60\,\Rsun$ and an
effective temperature of $\Teff=8000\,\K$. The temperature of the CSE
is not prescribed but computed from the~\citet{lucy1999} mean
intensity estimator. This choice of $\Teff$ gives at the inner radius
of model~b a dust temperature of $\simeq 1150\,\K$ lower than the
sublimation temperature. In this way, we can test if in the fitting
process using FRACS, a spurious effect might not lead the minimisation
algorithm to reach the upper limit for $\Tin$ of $1500\,\K$,
corresponding to the adopted dust sublimation temperature.


We obtained the best fitting parameters for the CSE model described in
Sect.~\ref{sect:testcase} with FRACS. For a comparison with the MC
code, $\alpha$ has been set and fixed to $4$ corresponding to the
value of a blackbody. Depending of the disc optical depth, the
temperature structure may show two separate regimes corresponding to
(1) the inner regions with the strongest temperature gradient,
optically thick to the stellar radiation and (2) the outer regions
optically thin to the disc radiation with a flatter temperature
gradient. In order to determine if mid-IR interferometric data are
sensitive to two temperature regimes, we tested the effect of two
parameterisations of the temperature structure: the unique power-law
of Eq.~(\ref{eq:temperature}) and a generalisation to two power-laws
with a transition radius, $\RT$, and a second exponent $\gamma'$
\begin{equation}
  \label{eq:temperature2}
  T(r)=\Tin\,\left(\frac{\Rin}{\RT}\right)^{\gamma}\times
  \left(\frac{\RT}{r}\right)^{\gamma'} \ ,
\end{equation}
for $r \ge \RT$. 

\begin{figure*}[!t]
  \centering
  $$
  \begin{array}{ccc}
    \resizebox{80mm}{!}{\includegraphics{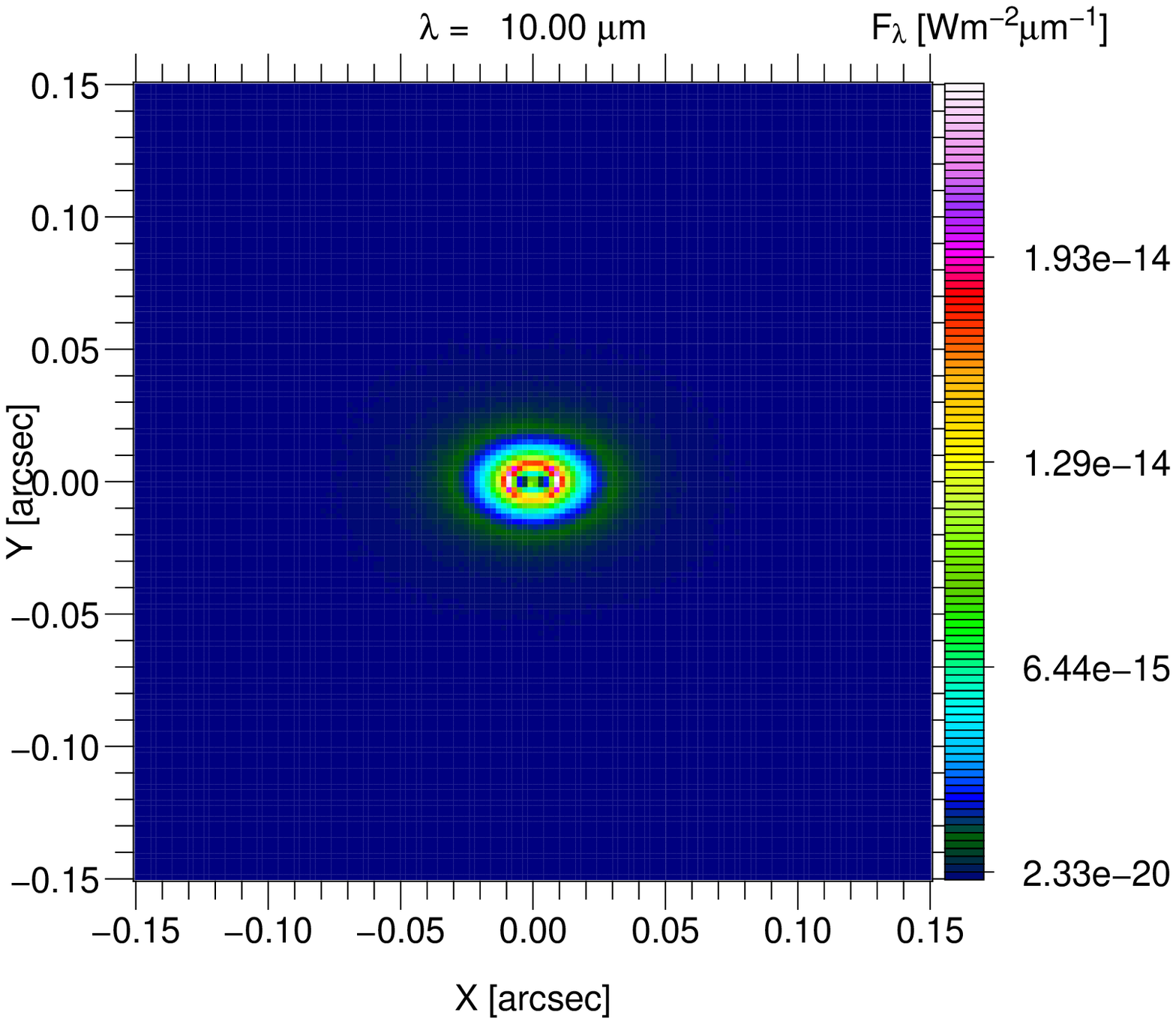}} &  {\hspace{.5cm}} &
    \resizebox{80mm}{!}{\includegraphics{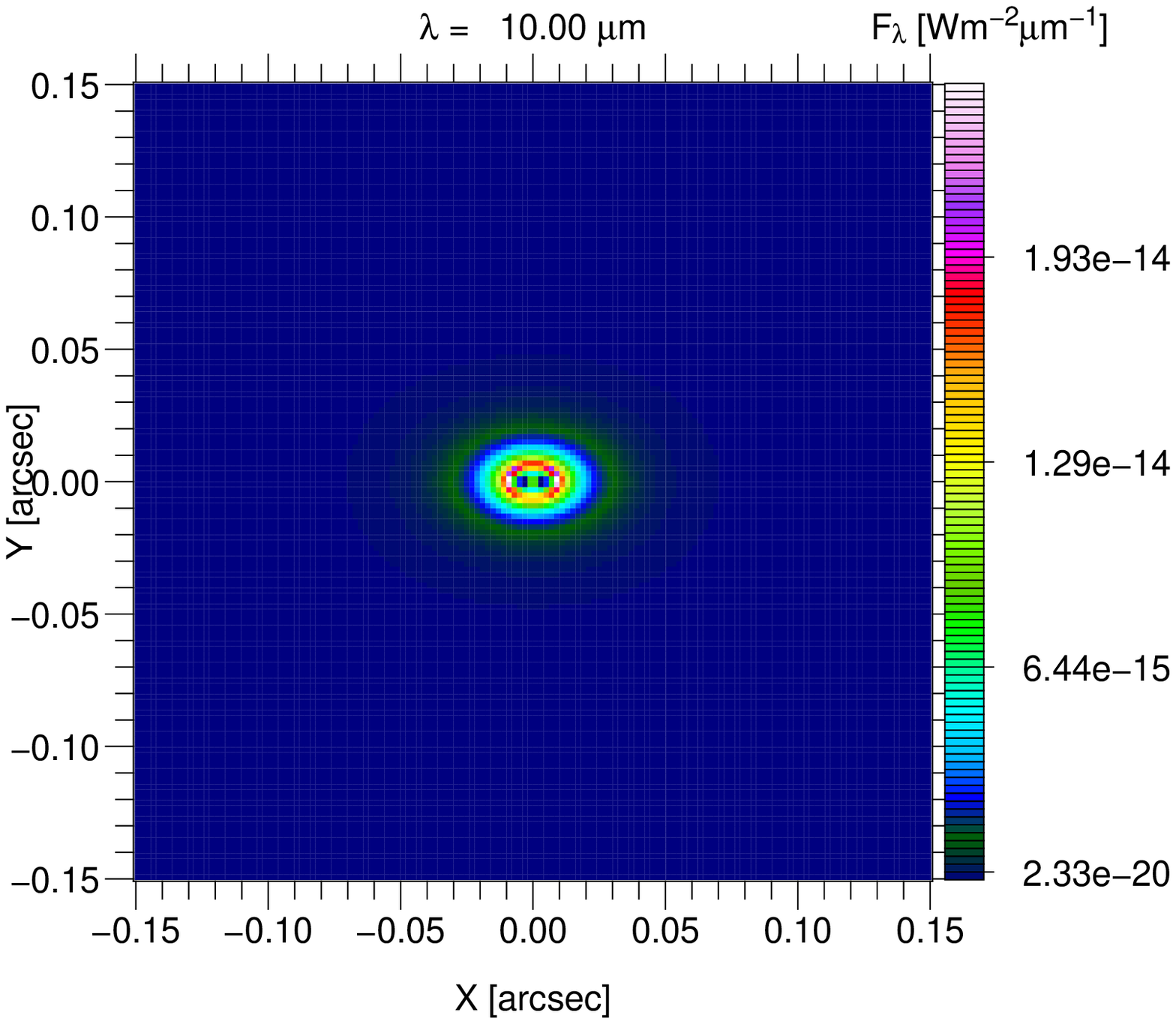}} \\
    \text{(a)} & {\hspace{.5cm}} & \text{(b)}
  \end{array}
  $$
  \caption{Disc images at $10\,\microns$. (a) Image computed with the
    help of the Monte Carlo radiative transfer code. (b) Image of the
    best-fitting model with two power-laws (parameters of the fourth
    column in Table.~\ref{tab:fracsvstapas}) obtained with FRACS.}
  \label{fig:images}
\end{figure*}

The best-fitting parameters for both parameterisations are shown in
Table~\ref{tab:fracsvstapas}. The images of the disc at $10\,\microns$
generated with the MC code and their corresponding FRACS counterpart
(best-fitting model) are shown in Fig.~\ref{fig:images} for
comparison.

\section{Discussion}
\label{sect:discussion}

We first discuss the uncertainties in the parameters derived for the
12~models studied in Sect.~\ref{sect:models}. For each model we
divided the parameters into three groups associated to a given level
of constraints expressed by the relative errors: below $10\,\%$,
between $10\,\%$ and $25\,\%$, and above $25\,\%$. This information is
summarised in Table~\ref{tab:res}. The exact relative errors for the
two models studied in detail (models~a and~b) are shown in
Table~\ref{tab:errors}.  Then, we analyse the results of
Sect.~\ref{sect:tapas} obtained from the best-fit of the data
simulated with the MC radiative transfer code.

\subsection{Central source}

Table~\ref{tab:res} shows that the central source parameters ($\phis$
and $\alpha$) can only be constrained with relative uncertainties
$\geq10\,\%$ for all test models. A deeper and more quantitative
investigation of these parameters can be obtained from models~(a)
and~(b). From Table~\ref{tab:errors} it can be seen that $\phis$ and
$\alpha$ are much better constrained for model~(a) ($27\,\%$ and
$13\,\%$, resp.) than for model~(b)(relative errors $\simeq100\,\%$).

The key quantity for a good constraint for the central source
parameters ($\phis$ and $\alpha$) is simply the relative contribution
of the flux of the central source to the total flux of the object
(source and disc). Indeed, the models in Table~\ref{tab:errors} only
differ by this relative flux contribution of $5.3\,\%$ in model~(a)
($\tau=0.1$), while it is only $0.7\,\%$ in model~(b) ($\tau=1$).

Our analysis thus shows that interferometric data can constrain
$\phis$ and $\alpha$ with a relative precision of
$\simeq15\,\%-30\,\%$ even when the central source contributes to
(only) a few percent of the total mid-IR flux.

\subsection{Geometrical parameters}

The parameters $\PAd$, $i$, and $\Rin$ are those usually estimated
from simple geometrical models (e.g.  ellipses, Gaussians). However,
their determination from geometrical models is quite limited, in
particular for $i$, for which only an estimate can be derived from the
axis-ratio of an ellipse, for example. In addition, the estimate of
$i$ from a simple analytical model such as a flat ellipse is only
valid for configurations far from the equator (intermediate to low
$i$). The use of a more physical and geometrically consistent model
such as FRACS allows us to relax this constraint and makes the
determination of $i$ possible for all viewing configurations.

\begin{figure*}[!t]
  \centering
  $$
  \begin{array}{cc}
    \resizebox{80mm}{!}{\includegraphics{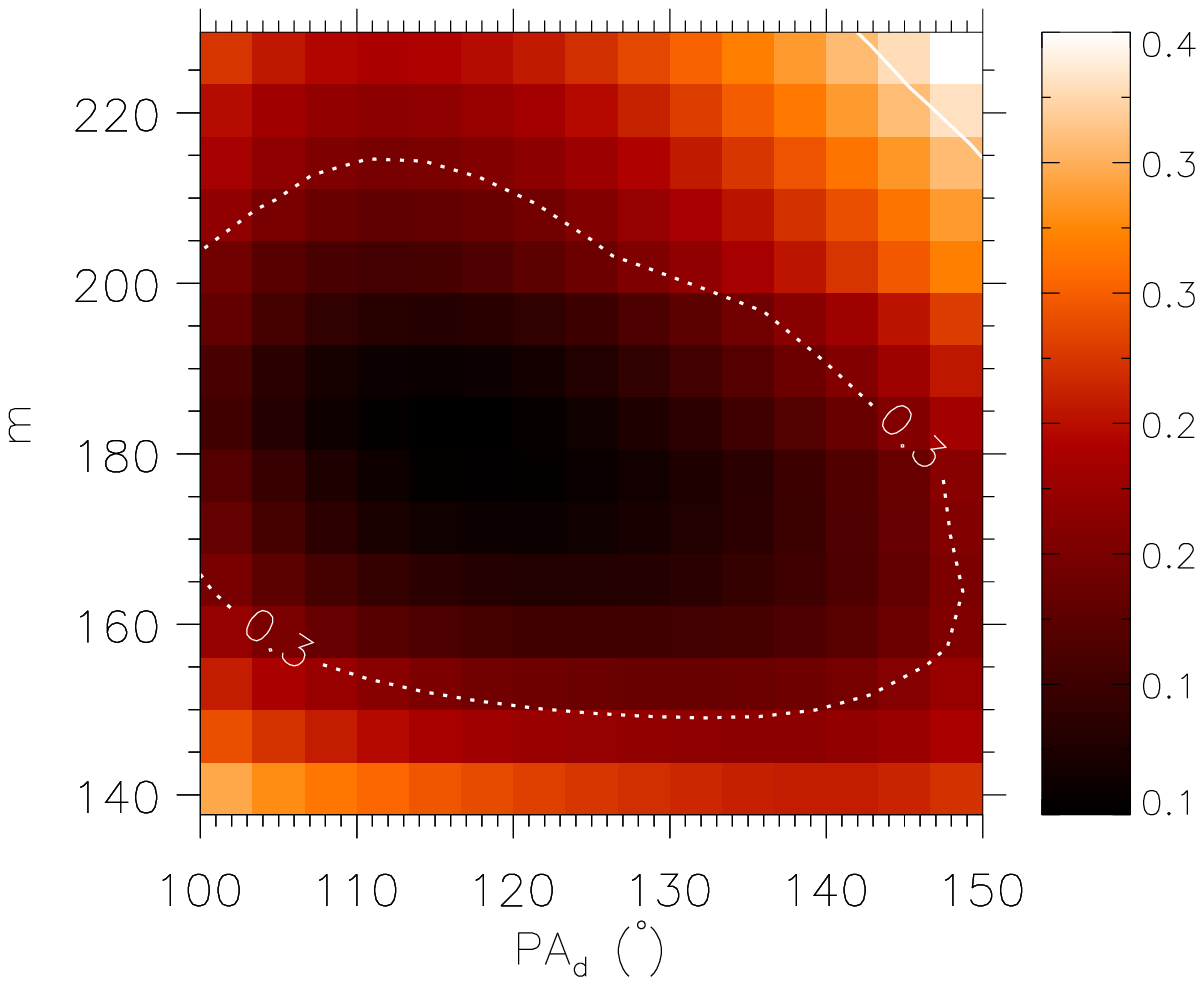}} &
    \resizebox{80mm}{!}{\includegraphics{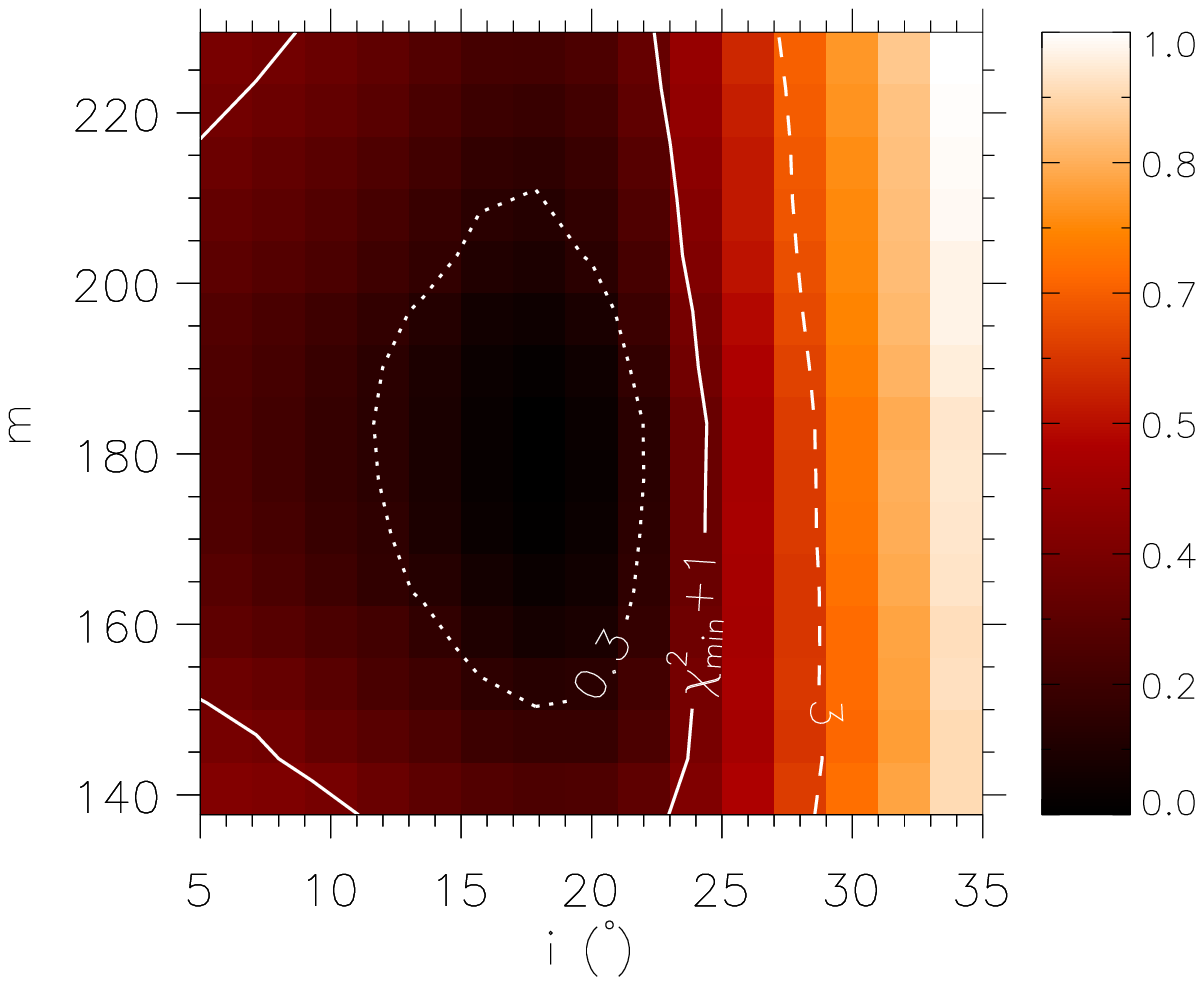}} \\
    \resizebox{80mm}{!}{\includegraphics{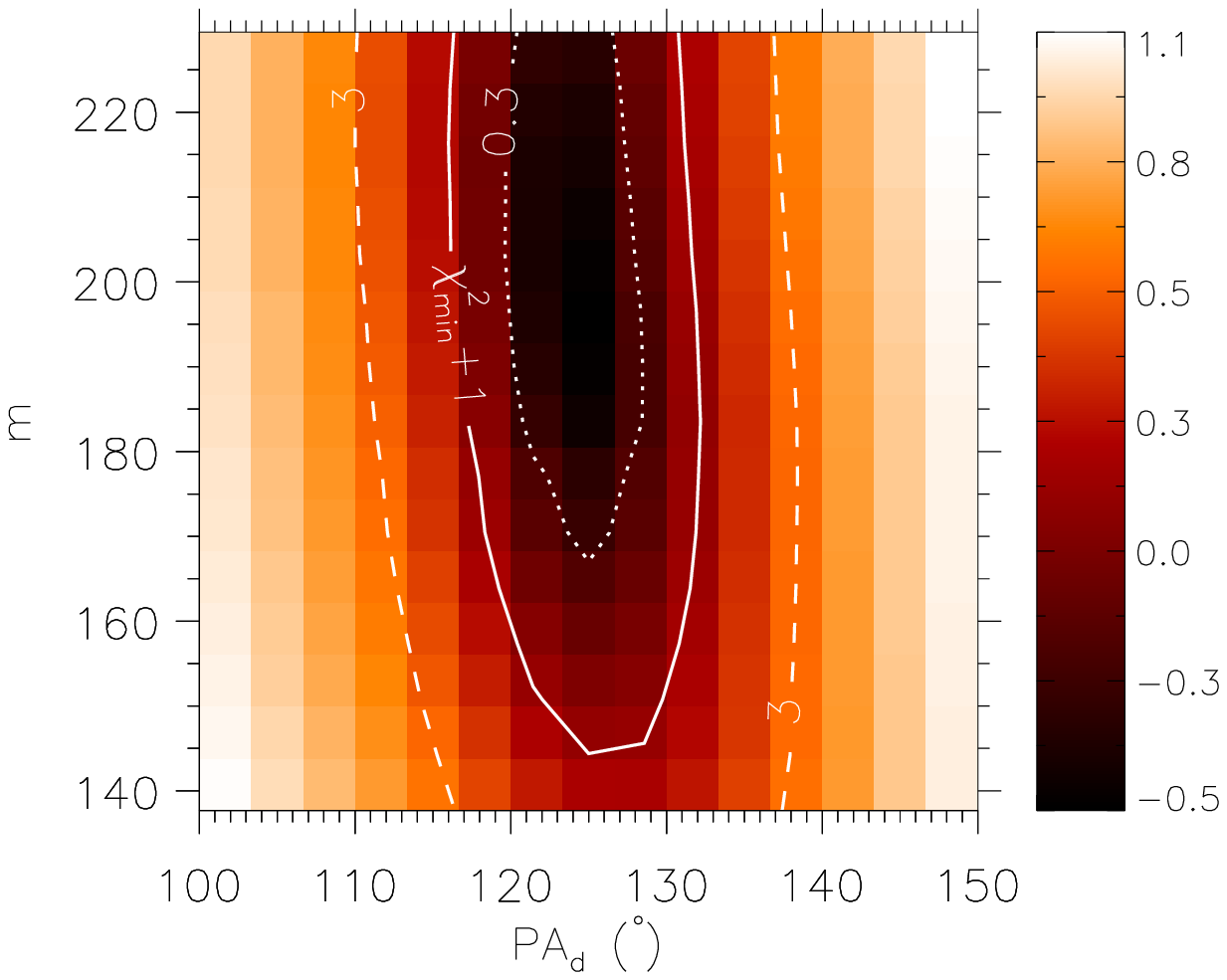}} &
    \resizebox{80mm}{!}{\includegraphics{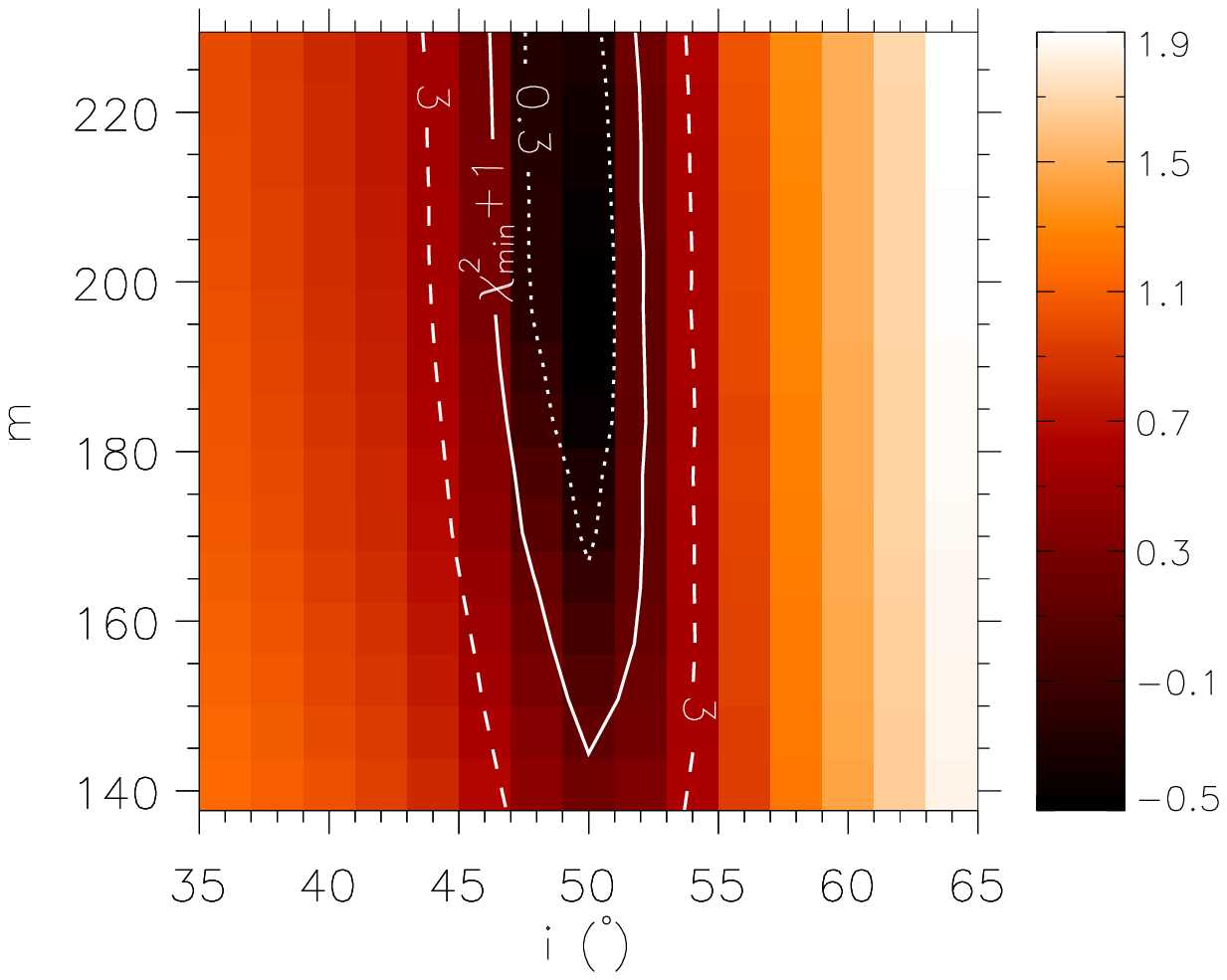}} \\
    \resizebox{80mm}{!}{\includegraphics{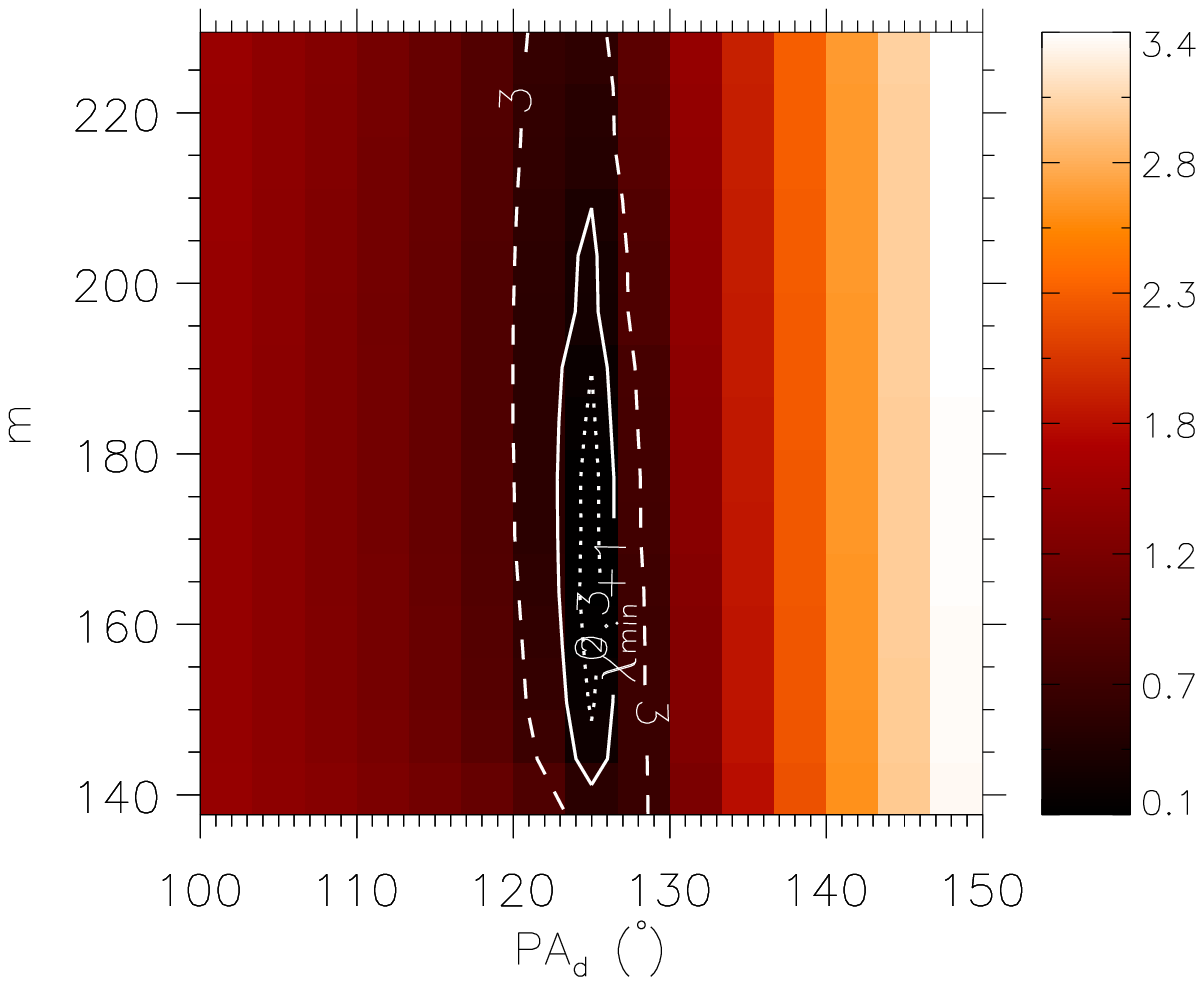}} &
    \resizebox{80mm}{!}{\includegraphics{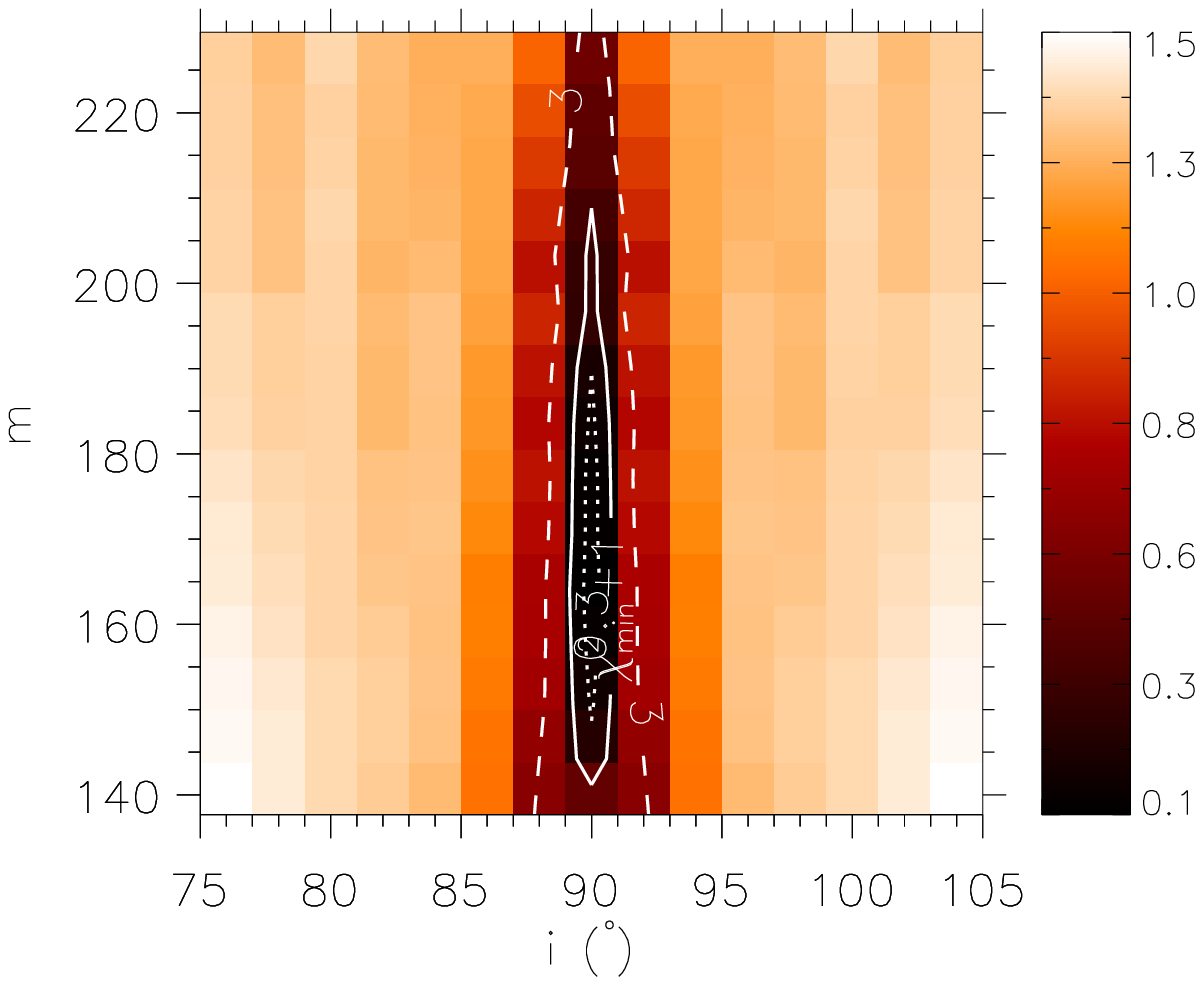}} \\
  \end{array}
  $$
  \caption{Evolution of $m$ and $\PAd$ with the inclination $i$.
    Left: $\chir^2$ maps for the couple $m$ and $\PAd$ ; right:
    $\chi^2$ maps for the couple $m$ and $i$. Contours are drawn for
    $\chirmin^2+\Delta \chir^2$, with $\Delta \chir^2=0.3,\,1,\,3$.
    From top to bottom the inclination $i$ takes the value
    $20\,\degr$, $50\,\degr$ and $90\,\degr$. The results correspond
    to model 4, 6, and 10. The limits of the maps have been set to
    $\pm 25\,\%$ of the true values of the parameters.}
  \label{fig:thetampa}
\end{figure*}

As expected, $\PAd$ and $i$ are better determined if the inclination
of the disc with respect to the line of sight is away from pole-on
(high $i$). In Fig.~\ref{fig:thetampa} we can clearly see this
behaviour from the $\chir^2$ maps involving $\PAd$ and $i$. Moreover,
the uncertainties on $\PAd$ and $i$ do not seem to be strongly
dependent on $\tau$ (equivalently $\rhoin$) and $\dthetadisc$
(equivalently $m$) for all models.

The inner dust radius $\Rin$ is not strongly dependent on any
parameter ($\tau$, $\dthetadisc$ or $i$), being very well constrained
(better than $10\,\%$) for most tested models.

\subsection{Temperature}

The parameters related to the temperature structure of the CSE, $\Tin$
and $\gamma$, are well constrained in most models, with relative
errors below $20\,\%$ and $12\,\%$ for both models~(a)
and~(b). Indeed, $\gamma$ has a strong impact on the IR emission
across the disc, and consequently this parameter has a direct
influence on the visibilities~(see Figs.~\ref{fig:visib1},
\ref{fig:visib2}).  $\Tin$ has a lower influence, compared to
$\gamma$, on the shape of the monochromatic image (radial dependence
of intensity) and can be mainly considered as a scaling factor to it.
On the other hand, the mid-IR flux imposes stronger constraints on
$\Tin$. From Table~\ref{tab:res} we see that the CSE's temperature
structure is not highly dependent on $\tau$ ($\rhoin$) and
$\dthetadisc$ for tested models.

\subsection{Number density of dust grains}

The parameters related to the density law, that is to say $m$,
$\rhoin$ and $A_2$, seems to be rather poorly constrained from the
mid-IR data alone. From the results of model~(a) and~(b) corresponding
to an intermediary inclination $i=50\,\degr$, we found that only
$\rhoin$ is constrained somewhat moderately with a mean relative error
of $46\,\%$.  For $m$ and $A_2$, according to the results of
Table~\ref{tab:errors} it seems that nevertheless, upper limits to
their values can be determined. Note that because $A_2$ is bounded
($-1 \le A_2 \le 0$) the mean relative errors, $53\,\%$ and $59\,\%$
for model~(a) and~(b) respectively, correspond approximatively to the
limit values of $A_2$, which is consequently not constrained.

Table~\ref{tab:res} confirms this trend for $m$, $\rhoin$ and $A_2$ at
least for the situations explored via the models presented here. From
all maps computed within $\pm 25\,\%$ of the true value, we always
found that the mean relative error to these parameters is larger than
$25\,\%$ with no hint that it could be close to these limits.

\begin{figure*}[!t]
  \centering
  $$
  \begin{array}{cc}
    \resizebox{80mm}{!}{\includegraphics{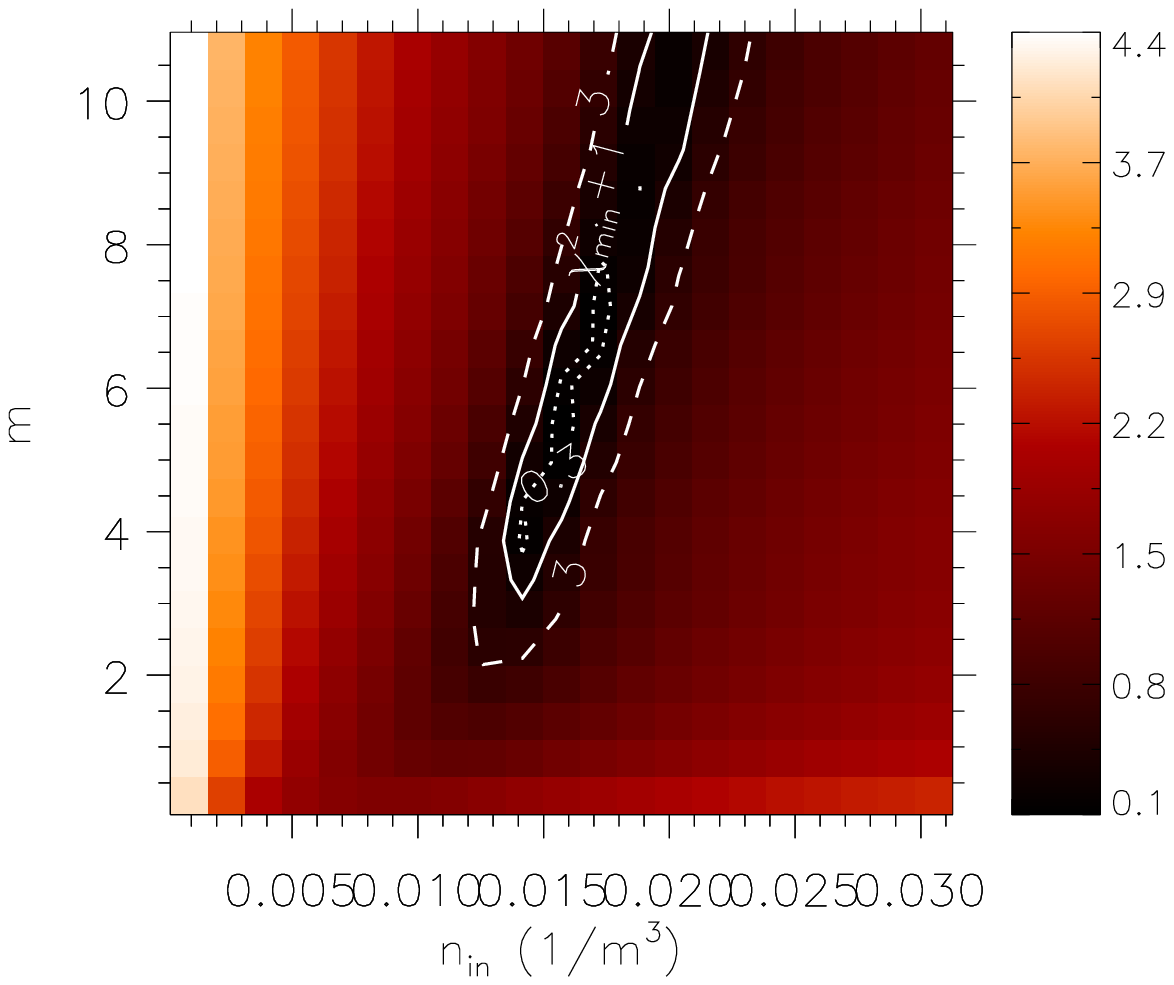}} &
    \resizebox{80mm}{!}{\includegraphics{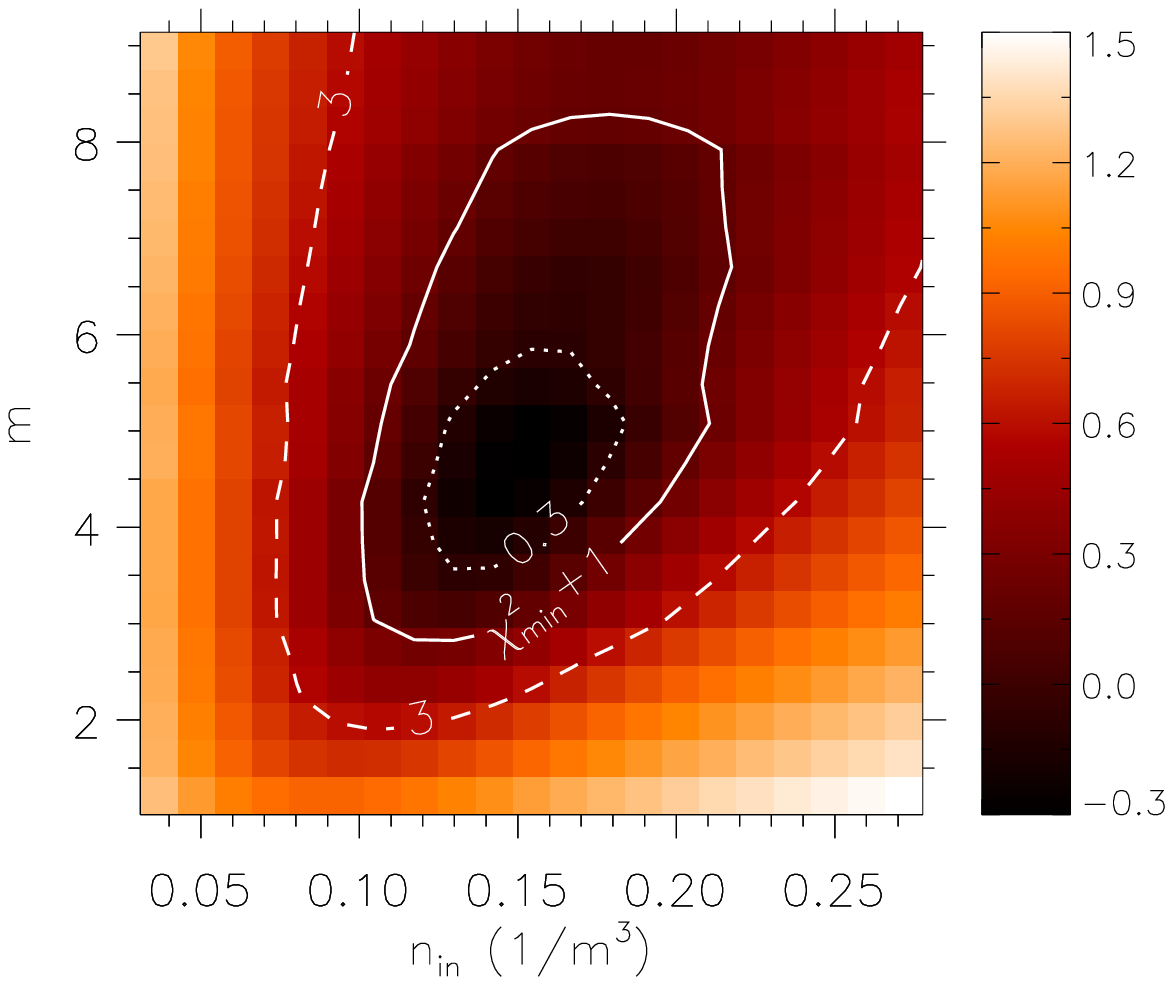}} \\
    \resizebox{80mm}{!}{\includegraphics{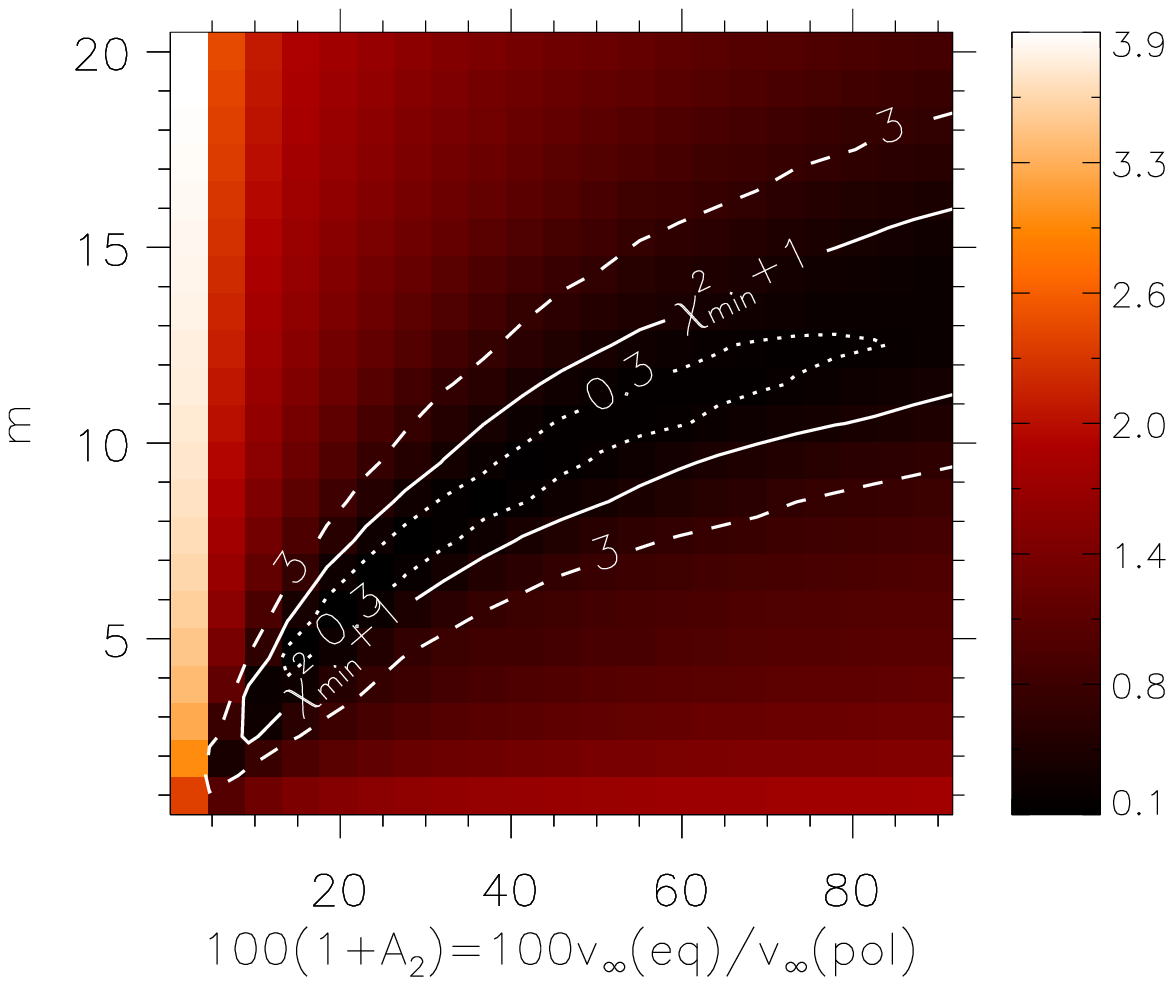}} &
    \resizebox{80mm}{!}{\includegraphics{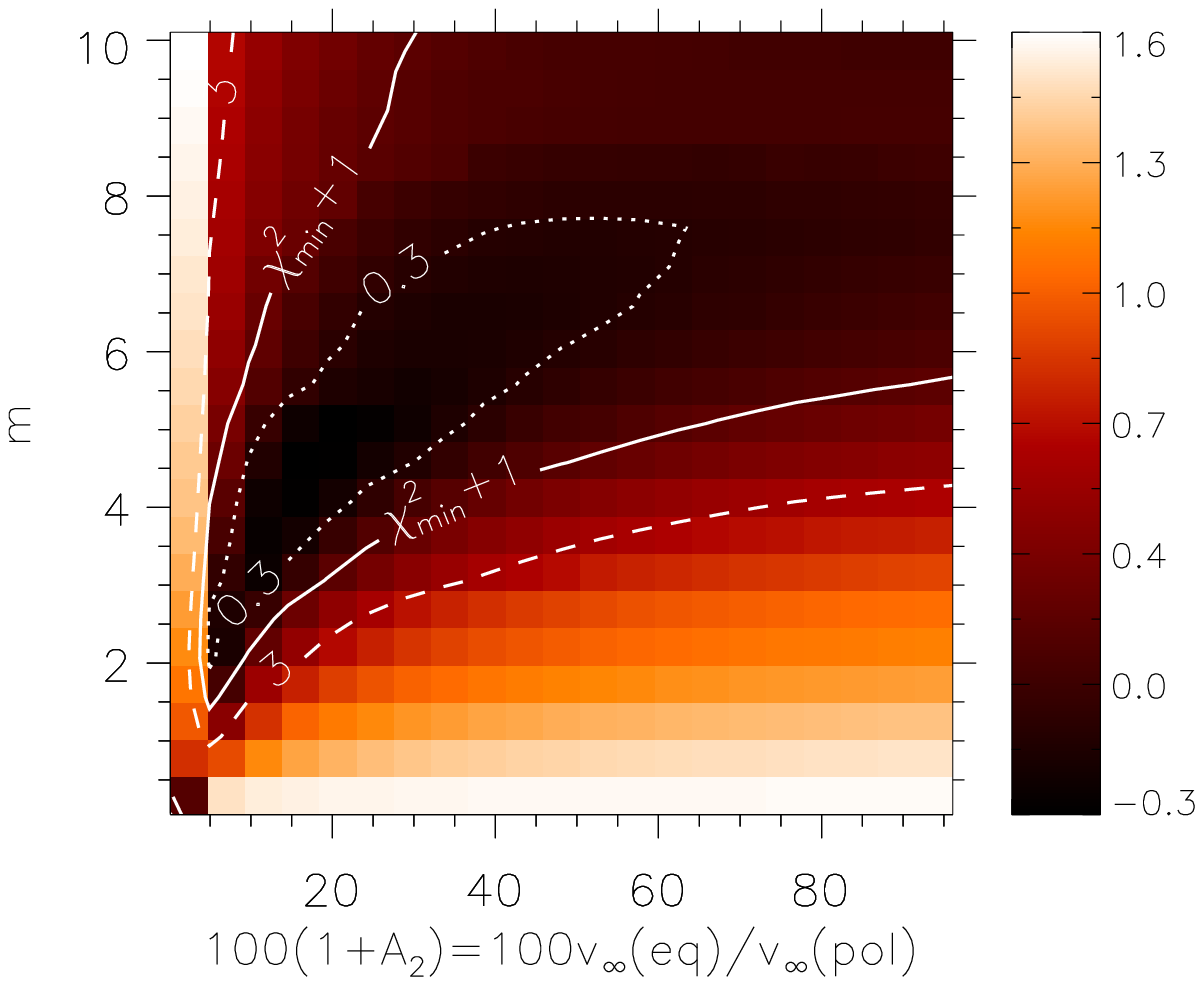}} \\
    \resizebox{80mm}{!}{\includegraphics{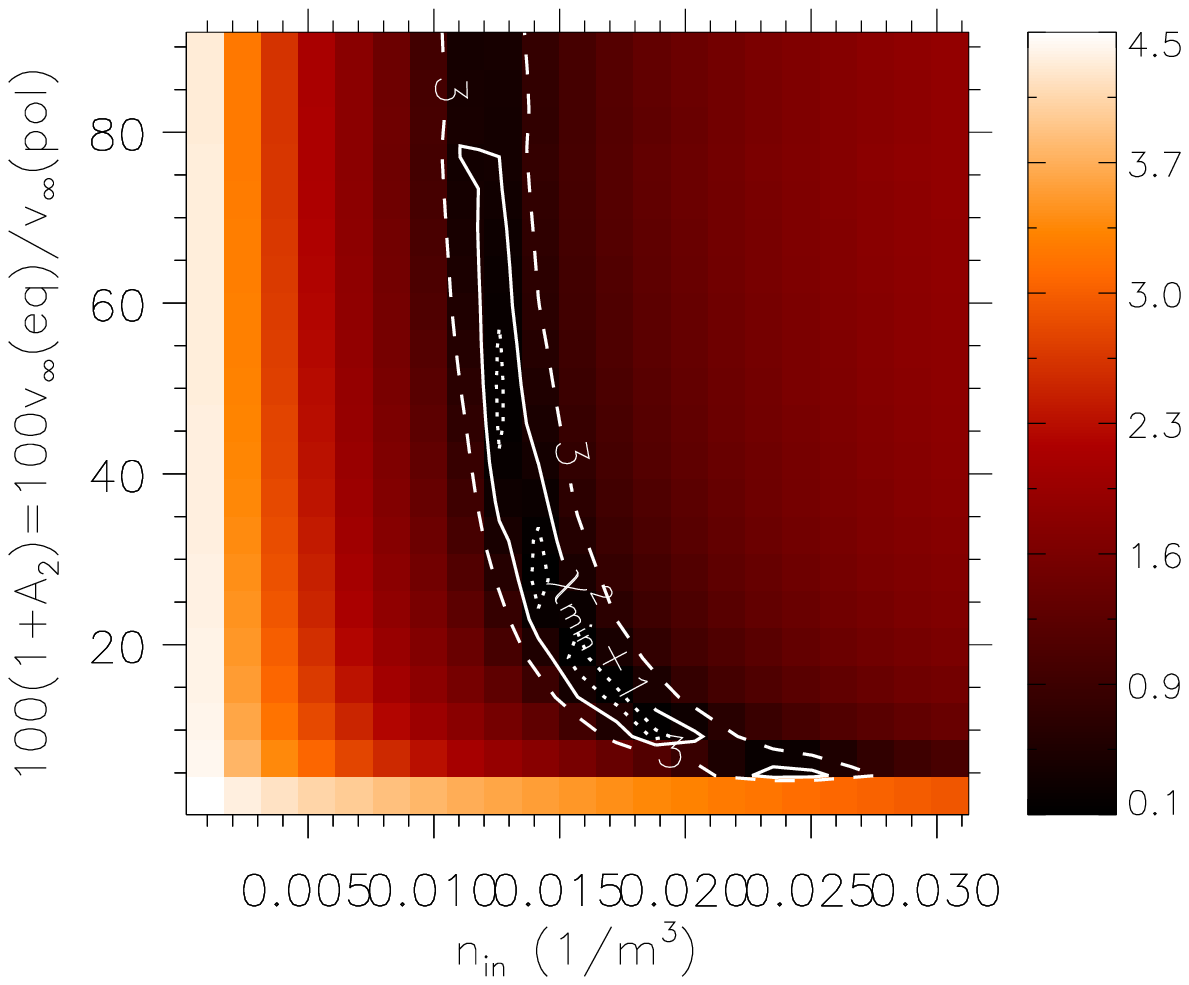}} &
    \resizebox{80mm}{!}{\includegraphics{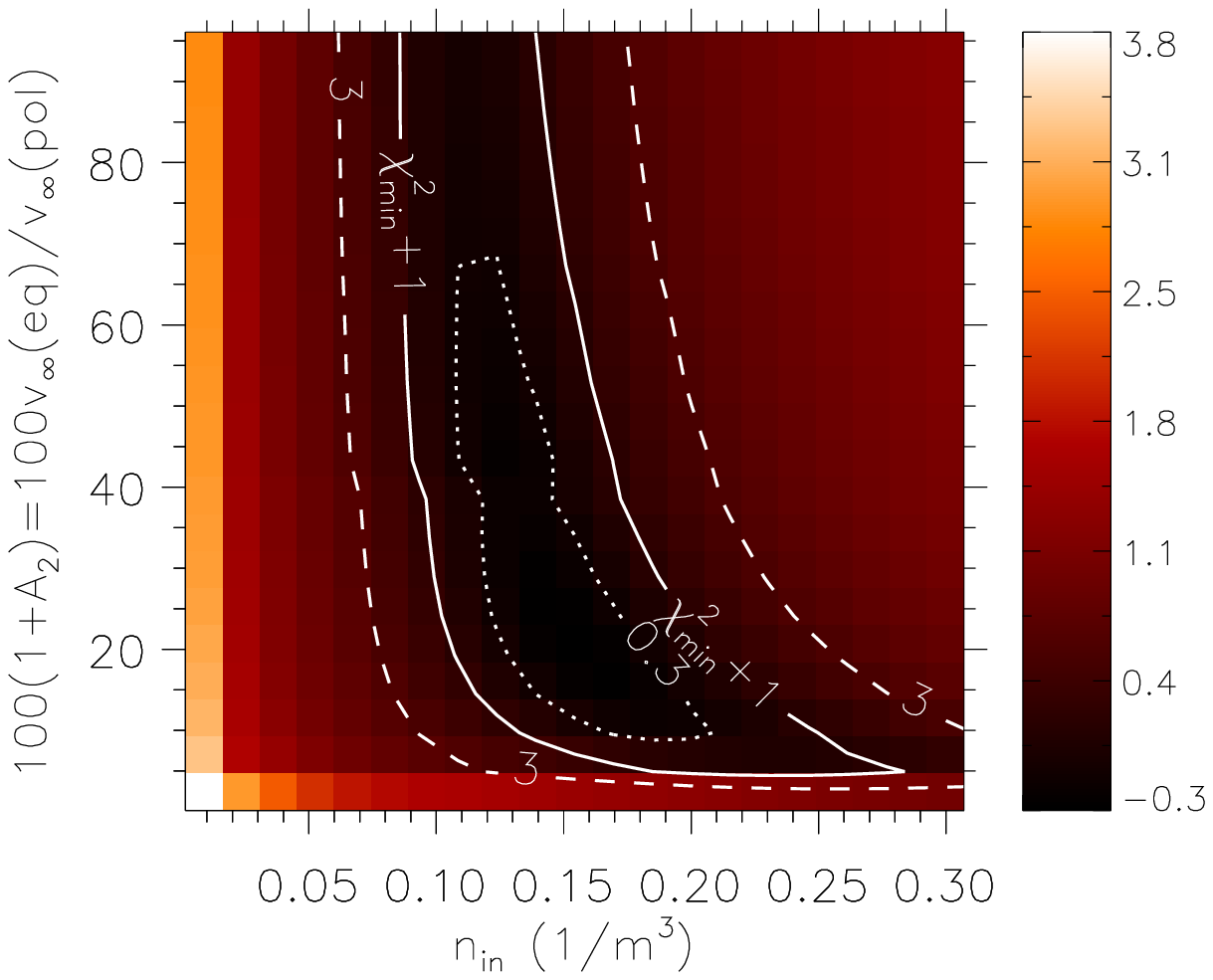}} \\
  \end{array}
  $$
  \caption{$\chir^2$ maps for the parameters $\rhoin$, $A_2$ and $m$.
    The results presented here are those of model~(a) ($\tau=0.1$,
    left part) and model~(b) ($\tau=1$, right part). Contours are
    drawn for $\chirmin^2+\Delta \chir^2$, with $\Delta
    \chir^2=0.3,\,1,\,3$. The three possible maps corresponding to the
    combination of these parameters are represented. These three
    parameters are better constrained in model~(a).}
  \label{fig:rho0A2m}
\end{figure*}

From Fig.~\ref{fig:rho0A2m}, comparing the $\chir^2$ maps for all
pairs of $\rhoin$, $m$ and $A_2$, we can see that the $\Delta \chir^2$
contours get sharper around the minimum value for model~(a)
(corresponding to lower optical depths along the line of sight) than
for model~(b). Indeed, the constraints on $\rhoin$ and $m$ are
improved for lower optical depths, or equivalently for lower disc
masses. Indeed, when the disc mass (or optical depth) decreases, the
flux (mid-IR flux, intensity maps) emitted by the disc reflects the
mass of the disc, while for high optical depths we only probe the
regions of the disc very close to the projected surface revealed to
the observer. $A_2$, however, is unaffected by the change in disc mass
and remains undetermined anyway. From Fig.~\ref{fig:rho0A2m} it can be
seen that $\rhoin$, $m$ and $A_2$ are strongly correlated. This is
expected from the expression of the density (see Eq.~\ref{eq:density})
depending on these parameters. However, this dependence and the final
correlation between these parameters are related through the
computation of the visibilities and the mid-IR flux, as well as the
comparison to the data and is, therefore, not straightforward.


To improve the situation concerning $\rhoin$, $m$ and $A_2$, the
mid-IR data can be supplemented by other types of observations such as
for instance spectroscopic data, from which one can better determine
$A_2$ \citep[\eg see ][]{chesneau2005}. We tested the effect of fixing
the value of $A_2$, or equivalently of assuming that $A_2$ is fully
determined, in the process of estimating the errors of the other
parameters. For model~(a), the relative errors on $\rhoin$ and $m$ go
down to $33\,\%$ and $71\,\%$ respectively while for model~(b),
$\rhoin$ and $m$ are determined with an accuracy reaching $95\,\%$ and
$78\,\%$ respectively. The precision to which other parameters are
determined is not affected by the determination of $A_2$.


We also tested the influence of the determination of $\rhoin$, $m$ and
$A_2$ on other parameters by fixing their values and estimatng the
relative errors on the remaining parameters for model~(a) and~(b).
Only $\phis$ and $\Tin$ are more strongly affected by the
determination of $\rhoin$, $m$ and $A_2$: for model~(a) (resp.
model~b) $\phis$ gets determined down to $19\,\%$ (resp. $80\,\%$) and
$\Tin$ down to $9\,\%$ (resp. $18\,\%$). The influence is stronger
with lower disc mass (model~a compared to model~b). This effect can be
explained because if we have a good determination of the disc mass
because we know $\rhoin$, $m$, and $A_2$, the determination of the
parameters that scale the source and disc fluxes is improved
accordingly for the visibility and the mid-IR flux.


$\rhoin$, $m$ and $A_2$ shape the density structure of the
circumstellar medium. Though they are not well constrained, they
certainly have a strong influence on the temperature structure, which
in turn is very well constrained. For the particular case of sgB[e]
circumstellar discs, a natural evolution of FRACS is to include the
direct heating of the medium by the central source of radiation
assuming that the disc is optically thin to its own radiation. The
temperature structure would not be parameterised, and its good
determination would certainly put better constraints on $\rhoin$, $m$
and $A_2$, while keeping an affordable computation time for the
model-fitting procedure. This will be the purpose of a subsequent
work.

Finally, one can derive a ranking of the parameter constraints
according to two criteria: first the parameter must be constrained
within the prescribed limits ($100\,\%$ for model a and~b and $25\,\%$
for model~1 to~10) and second the mean relative error must be as low
as possible. The best-fitted parameters, most of the time according to
these criteraria are by decreasing order of best determination:
$\Rin$, $\PAd$, $\gamma$, $\Tin$, $i$, $\alpha$, $\phis$, $\rhoin$,
$m$ and $A_2$. This tendency can be seen in Table~\ref{tab:res}.


\subsection{best-fit to the MC simulation}


\begin{figure}[t!]
  \centering
  \resizebox{8.cm}{!}{\includegraphics{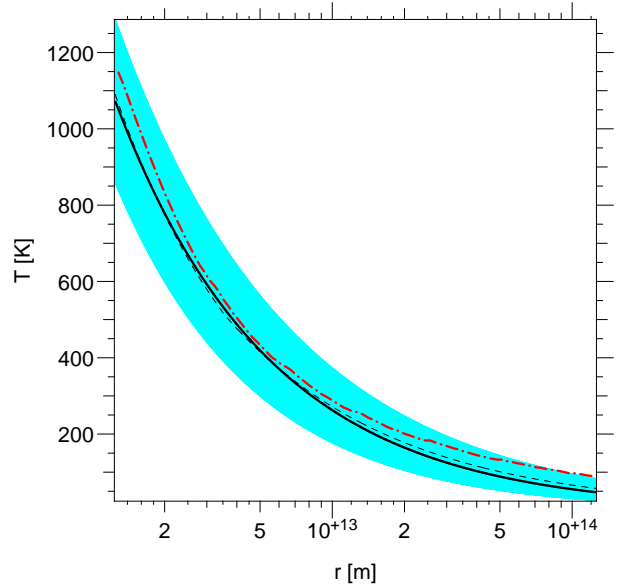}}
  \caption{Temperature of the CSE. The solid line represents the best
    fit (last column in Table.~\ref{tab:fracsvstapas}) with a unique
    power-law, the dashed line the best-fit with two power-laws
    (fourth column in Table~\ref{tab:fracsvstapas}) and the dot-dashed
    line the MC results. The shaded region represents the possible
    domain for a unique power-law by taking into account the errors
    estimated in Table~\ref{tab:errors}.}
  \label{fig:temp}
\end{figure}

The $\chirmin^2$ values obtained for the two types of temperature
parametrisations (one and two power-laws) are quite similar: $0.79$
and $0.73$ respectively. Regarding the data, both temperature
parametrisations are indeed acceptable. In addition, these results
show that we can actually obtain very good fits from data sets based
on more physically consistent scenarios. A complete error analysis and
study of the parameter determination has been presented in the
previous sections for data generated from FRACS and will not be
repeated here. In particular, parameter confidence intervals, from
which errors were derived, have already been estimated. Here, we will
instead focus on the {\em true} errors, \ie the differences between
the true model parameters and the best-fitting values for the
parameters (see Table~\ref{tab:fracsvstapas}). The two types of errors
must not be confounded. The {\em true} errors reflect the capability
of FRACS to mimic the mid-IR interferometric data regarding the
information it provides. Of course, with the sparse uv-plane coverage
inherent to this kind of data as well as the experimental noise, one
should not expect a full agreement of the fitted and the true
parameters: they are indeed different.


From Table~\ref{tab:fracsvstapas}, we see that the geometrical
parameters, $\PAd$, $i$ and $\Rin$, can be almost exactly recovered as
expected. The source specific intensity $\phis$, and the parameters
related to the density, $A_2$, $\rhoin$ and $m$ can be recovered
fairly well and have best-fitting values close to the true parameters.


The values of $\Tin$, $\RT$, $\gamma$ and $\gamma'$ reported as
``true'' in Table~\ref{tab:fracsvstapas} are indeed the values of a
fit to the average (over the co-latitude for a given $r$) {\em
  computed} temperature in the disc. The true relative differences for
$\Tin$ do not exceed $7\,\%$ independently of the adopted
parameterisation of the temperature (one or two power-laws). The
best-fitting values of $\gamma$, the inner temperature gradient,
obtained with FRACS are very close to the true values with two and one
power-law with true relative error of $1\,\%$ and $7\,\%$
respectively. This already suggests that the mid-IR data provide
information on the inner and {\em hottest} region of the CSE, in
particular on the inner temperature gradient $\gamma$.


Fitting the temperature computed with the MC code with a simple
power-law, we obtain $\gamma \simeq 0.64$. This value is close to
those of the best fitting models, especially with a unique power-law
($6\,\%$ relative difference). For comparison, the actual mean
temperature gradient as derived from the MC simulation is $\simeq
0.60$. For this particular data set, the values of $\gamma'$ and $\RT$
recovered by FRACS differ by $28\,\%$ and $45\,\%$ respectively from
the actual values. This again confirms the sensibility of the
interferometric data to the temperature structure mostly in the inner
($r \la \RT$) regions of the disc. The best-fitting models (fourth and
last columns in Table~\ref{tab:fracsvstapas}) as well as the MC
results are shown in Fig.~\ref{fig:temp}. Regarding the errors
(estimated from the results given in Table~\ref{tab:errors}) shown as
a shaded area, we can see that both temperature parameterisations are
essentially the same and show a better agreement with the MC results
in the inner than in the outer regions of the disc.

We considered a ``truncated'' model with two power-laws (with
parameter values listed in the third column of
Table~\ref{tab:fracsvstapas}) in which the CSEs emission for $r \ge
\RT$, the ``outer'' regions, has been set to 0. We then compared the
visibilities and the fluxes of this truncated model to the same model
{\em including} the outer region emission. We obtained relative
differences, averaged over all considered wavelengths and baselines
(Table~\ref{tab:BPA}), of $18\,\%$ and $17\,\%$ respectively. These
relative differences are larger, but are still close to the noise
level. For this reason, one cannot expect to obtain much information
on the outer temperature gradient $\gamma'$, at least for the
particular configuration we considered.




\section{Conclusion}
\label{sect:conclusion}



We proposed and described here a new numerical tool to interpret
mid-IR interferometric data. Even though we focussed on the special
case of circumstellar disc observations, the numerical techniques have
been developed with the aim to be as general as possible. The methods
we employ rely on both parameterised physical models and the
ray-tracing technique. The need for such a tool is evident because the
nature of interferometric data imposes an interpretation through a
model of the object to obtain any kind of information. On one hand,
Monte-Carlo radiative transfer methods require too much computation
time to associate the model-fitting to an automatic minimum search
method. On the other hand, purely geometrical function fitting (such
as ellipses or Gaussians) are too simple to envisage to obtain
physical constraints on the observed disc. Hence, a tool like FRACS
fills a blank in the model fitting approach for mid-IR interferometric
data interpretation. The main advantages of FRACS are its speed and
its flexibility, allowing us to test different physical
models. Moreover, an exploration of the parameter space can be
performed in different manners and can lead to an estimate of the
sensitivity of the fit to the different model parameters, \ie a
realistic error estimate.

We applied these techniques to the special astrophysical case of B[e]
star circumstellar environments by generating artificial data in order
to analyse beforehand what constraints can be obtained on each
parameter of the particular disc model in this work. The techniques
will then be applied to real interferometric data of a sgB[e] CSE in a
sequel to this paper.


We showed in our analysis that the ``geometrical'' parameters such as
$\Rin$, $\PAd$ and $i$ can be determined with an accuracy $\la
15\,\%$. Mid-IR interferometric data give access to a mean temperature
gradient: the temperature structure ($\Tin$ and $\gamma$) can be very
well determined (within $\la 20\,\%$ and $\la 10\,\%$
respectively). It is possible to have access to the central source
emission (with an accuracy $\ga 30 \,\%$) when it has a significant
contribution to the total flux of the object (a few $\%$ are
sufficient). The remaining parameters of our disc model, namely
$\rhoin$, $m$ and $A_2$ are not very well constrained by MIDI data
alone. $\rhoin$ is at best determined with an accuracy of about $\ga
50\,\%$ in some cases. If $A_2$ can be estimated through spectroscopic
observations, then the picture about the $\rhoin$ and $m$
determination improves somewhat.



FRACS can be used mainly for two purposes. First, it can be used by
itself to try and determine physical quantities of the circumstellar
matter. Admittedly, it is not a self-consistent model, \ie the
radiative transfer is not solved because the temperature structure is
parameterised. From the usual habits in the interpretation of
interferometric data it is nevertheless a step beyond the commonly use
of toy models or very simple analytical models. This approach has
indeed been very successful in the millimetric wavelength
range~\citep[\eg see][]{guilloteau1998}. Second, it can be viewed as a
mean to prepare the work of data fitting with a more elaborate model
(such as a Monte Carlo radiative transfer code for instance) and to
provide a good starting point.


FRACS is a tool that can help in the process of interpreting and/or
preparing observations with second-generation VLTI instruments such as
the Multi-AperTure mid-Infrared SpectroScopic Experiment (MATISSE)
project~\citep{lopez2006}. In this respect, FRACS is not restricted to
the mid-IR, and sub-millimeter interferometric data obtained with the
Atacama Large Millimeter Array (ALMA) for instance can be tackled.
  


\begin{acknowledgements}
  We thank the anonymous referee for her/his constructive comments
  that lead to significant improvements of the manuscript. We also
  would like to thank Alex Carciofi, Olga Su\'{a}rez Fern\'{a}ndez,
  Andrei and Ivan Belokogne for fruitful discussions and careful proof
  reading.

  The Monte Carlo simulation have been carried out on a computer
  financed by the BQR grant of the Observatoire de la C\^{o}te d'Azur.

  This work is dedicated to Lucien.
\end{acknowledgements}


\bibliographystyle{aa}
\bibliography{bibliography}

\begin{thebibliography}{35}
\expandafter\ifx\csname natexlab\endcsname\relax\def\natexlab#1{#1}\fi

\bibitem[{{Bianchi}(2008)}]{bianchi2008}
{Bianchi}, S. 2008, \aap, 490, 461

\bibitem[{{Carciofi} {et~al.}(2010){Carciofi}, {Miroshnichenko}, \&
  {Bjorkman}}]{carciofi2010}
{Carciofi}, A.~C., {Miroshnichenko}, A.~S., \& {Bjorkman}, J.~E. 2010, \apj,
  721, 1079

\bibitem[{{Chesneau} {et~al.}(2005){Chesneau}, {Meilland}, {Rivinius}, {Stee},
  {Jankov}, {Domiciano de Souza}, {Graser}, {Herbst}, {Janot-Pacheco},
  {Koehler}, {Leinert}, {Morel}, {Paresce}, {Richichi}, \&
  {Robbe-Dubois}}]{chesneau2005}
{Chesneau}, O., {Meilland}, A., {Rivinius}, T., {et~al.} 2005, \aap, 435, 275

\bibitem[{{di Folco} {et~al.}(2009){di Folco}, {Dutrey}, {Chesneau}, {Wolf},
  {Schegerer}, {Leinert}, \& {Lopez}}]{difolco2009}
{di Folco}, E., {Dutrey}, A., {Chesneau}, O., {et~al.} 2009, \aap, 500, 1065

\bibitem[{{Domiciano de Souza} {et~al.}(2007){Domiciano de Souza}, {Driebe},
  {Chesneau}, {Hofmann}, {Kraus}, {Miroshnichenko}, {Ohnaka}, {Petrov},
  {Preisbisch}, {Stee}, {Weigelt}, {Lisi}, {Malbet}, \&
  {Richichi}}]{domiciano2007}
{Domiciano de Souza}, A., {Driebe}, T., {Chesneau}, O., {et~al.} 2007, \aap,
  464, 81

\bibitem[{{Draine} \& {Lee}(1984)}]{draine1984}
{Draine}, B.~T. \& {Lee}, H.~M. 1984, \apj, 285, 89

\bibitem[{{Feiveson} \& {Delaney}(1968)}]{feiveson1968}
{Feiveson}, A.~H. \& {Delaney}, F.~C. 1968, {The distribution and properties of
  a weighted sum of chi squares}, Tech. rep., National Aeronautics and Space
  Administration

\bibitem[{{Felli} \& {Panagia}(1981)}]{felli1981}
{Felli}, M. \& {Panagia}, N. 1981, \aap, 102, 424

\bibitem[{{Guilloteau} \& {Dutrey}(1998)}]{guilloteau1998}
{Guilloteau}, S. \& {Dutrey}, A. 1998, \aap, 339, 467

\bibitem[{{Jonsson}(2006)}]{jonsson2006}
{Jonsson}, P. 2006, \mnras, 372, 2

\bibitem[{{Kurosawa} \& {Hillier}(2001)}]{kurosawa2001}
{Kurosawa}, R. \& {Hillier}, D.~J. 2001, \aap, 379, 336

\bibitem[{{Lachaume} {et~al.}(2007){Lachaume}, {Preibisch}, {Driebe}, \&
  {Weigelt}}]{lachaume2007}
{Lachaume}, R., {Preibisch}, T., {Driebe}, T., \& {Weigelt}, G. 2007, \aap,
  469, 587

\bibitem[{{Lamers} \& {Cassinelli}(1999)}]{lamers1999}
{Lamers}, H.~J.~G.~L.~M. \& {Cassinelli}, J.~P. 1999, {Introduction to Stellar
  Winds}, ed. J.~P. Lamers, H. J. G. L. M. \&~Cassinelli

\bibitem[{{Lamers} \& {Waters}(1987)}]{lamers1987}
{Lamers}, H.~J.~G.~L.~M. \& {Waters}, L.~B.~F.~M. 1987, \aap, 182, 80

\bibitem[{{Lamers} {et~al.}(1998){Lamers}, {Zickgraf}, {de Winter}, {Houziaux},
  \& {Zorec}}]{lamers1998}
{Lamers}, H.~J.~G.~L.~M., {Zickgraf}, F., {de Winter}, D., {Houziaux}, L., \&
  {Zorec}, J. 1998, \aap, 340, 117

\bibitem[{{Leinert} {et~al.}(2003){Leinert}, {Graser}, {Przygodda}, {Waters},
  {Perrin}, {Jaffe}, {Lopez}, {Bakker}, {B{\"o}hm}, {Chesneau}, {Cotton},
  {Damstra}, {de Jong}, {Glazenborg-Kluttig}, {Grimm}, {Hanenburg}, {Laun},
  {Lenzen}, {Ligori}, {Mathar}, {Meisner}, {Morel}, {Morr}, {Neumann}, {Pel},
  {Schuller}, {Rohloff}, {Stecklum}, {Storz}, {von der L{\"u}he}, \&
  {Wagner}}]{leinert2003}
{Leinert}, C., {Graser}, U., {Przygodda}, F., {et~al.} 2003, \apss, 286, 73

\bibitem[{{Leinert} {et~al.}(2004){Leinert}, {van Boekel}, {Waters},
  {Chesneau}, {Malbet}, {K{\"o}hler}, {Jaffe}, {Ratzka}, {Dutrey}, {Preibisch},
  {Graser}, {Bakker}, {Chagnon}, {Cotton}, {Dominik}, {Dullemond},
  {Glazenborg-Kluttig}, {Glindemann}, {Henning}, {Hofmann}, {de Jong},
  {Lenzen}, {Ligori}, {Lopez}, {Meisner}, {Morel}, {Paresce}, {Pel},
  {Percheron}, {Perrin}, {Przygodda}, {Richichi}, {Sch{\"o}ller}, {Schuller},
  {Stecklum}, {van den Ancker}, {von der L{\"u}he}, \& {Weigelt}}]{leinert2004}
{Leinert}, C., {van Boekel}, R., {Waters}, L.~B.~F.~M., {et~al.} 2004, \aap,
  423, 537

\bibitem[{{Levenberg}(1944)}]{levenberg1944}
{Levenberg}, K. 1944, The Quarterly of Applied Mathematics, 2, 164

\bibitem[{{Lopez} {et~al.}(2006){Lopez}, {Wolf}, {Lagarde}, {Abraham},
  {Antonelli}, {Augereau}, {Beckman}, {Behrend}, {Berruyer}, {Bresson},
  {Chesneau}, {Clausse}, {Connot}, {Demyk}, {Danchi}, {Dugu{\'e}}, {Flament},
  {Glazenborg}, {Graser}, {Henning}, {Hofmann}, {Heininger}, {Hugues}, {Jaffe},
  {Jankov}, {Kraus}, {Laun}, {Leinert}, {Linz}, {Mathias}, {Meisenheimer},
  {Matter}, {Menut}, {Millour}, {Neumann}, {Nussbaum}, {Niedzielski},
  {Mosonic}, {Petrov}, {Ratzka}, {Robbe-Dubois}, {Roussel}, {Schertl},
  {Schmider}, {Stecklum}, {Thiebaut}, {Vakili}, {Wagner}, {Waters}, \&
  {Weigelt}}]{lopez2006}
{Lopez}, B., {Wolf}, S., {Lagarde}, S., {et~al.} 2006, in Society of
  Photo-Optical Instrumentation Engineers (SPIE) Conference Series, Vol. 6268,
  Society of Photo-Optical Instrumentation Engineers (SPIE) Conference Series

\bibitem[{{Lucy}(1999)}]{lucy1999}
{Lucy}, L.~B. 1999, \aap, 344, 282

\bibitem[{{Malbet} {et~al.}(2005){Malbet}, {Lachaume}, {Berger}, {Colavita},
  {di Folco}, {Eisner}, {Lane}, {Millan-Gabet}, {S{\'e}gransan}, \&
  {Traub}}]{malbet2005}
{Malbet}, F., {Lachaume}, R., {Berger}, J., {et~al.} 2005, \aap, 437, 627

\bibitem[{{Marquardt}(1963)}]{marquardt1963}
{Marquardt}, D. 1963, SIAM Journal on Applied Mathematics, 11, 431

\bibitem[{{Mathis} {et~al.}(1977){Mathis}, {Rumpl}, \&
  {Nordsieck}}]{mathis1977}
{Mathis}, J.~S., {Rumpl}, W., \& {Nordsieck}, K.~H. 1977, \apj, 217, 425

\bibitem[{{Mie}(1908)}]{mie1908}
{Mie}, G. 1908, \anphys, 25, 377

\bibitem[{{Niccolini} \& {Alcolea}(2006)}]{niccolini2006}
{Niccolini}, G. \& {Alcolea}, J. 2006, \aap, 456, 1

\bibitem[{{Ohnaka} {et~al.}(2006){Ohnaka}, {Driebe}, {Hofmann}, {Leinert},
  {Morel}, {Paresce}, {Preibisch}, {Richichi}, {Schertl}, {Sch{\"o}ller},
  {Waters}, {Weigelt}, \& {Wittkowski}}]{ohnaka2006}
{Ohnaka}, K., {Driebe}, T., {Hofmann}, K., {et~al.} 2006, \aap, 445, 1015

\bibitem[{{Panagia} \& {Felli}(1975)}]{panagia1975}
{Panagia}, N. \& {Felli}, M. 1975, \aap, 39, 1

\bibitem[{{Porter}(2003)}]{porter2003}
{Porter}, J.~M. 2003, \aap, 398, 631

\bibitem[{{Press} {et~al.}(1992){Press}, {Teukolsky}, {Vetterling}, \&
  {Flannery}}]{press1992}
{Press}, W.~H., {Teukolsky}, S.~A., {Vetterling}, W.~T., \& {Flannery}, B.~P.
  1992, {Numerical recipes in C. The art of scientific computing} (Cambridge:
  University Press, |c1992, 2nd ed.)

\bibitem[{{Stee} {et~al.}(1995){Stee}, {de Araujo}, {Vakili}, {Mourard},
  {Arnold}, {Bonneau}, {Morand}, \& {Tallon-Bosc}}]{stee1995}
{Stee}, P., {de Araujo}, F.~X., {Vakili}, F., {et~al.} 1995, \aap, 300, 219

\bibitem[{{Steinacker} {et~al.}(2006){Steinacker}, {Bacmann}, \&
  {Henning}}]{steinacker2006}
{Steinacker}, J., {Bacmann}, A., \& {Henning}, T. 2006, \apj, 645, 920

\bibitem[{{Wiscombe}(1980)}]{wiscombe1980}
{Wiscombe}, W.~J. 1980, \ao, 19, 1505

\bibitem[{{Wolf} {et~al.}(1999){Wolf}, {Henning}, \& {Stecklum}}]{wolf1999}
{Wolf}, S., {Henning}, T., \& {Stecklum}, B. 1999, \aap, 349, 839

\bibitem[{{Zickgraf}(2003)}]{zickgraf2003}
{Zickgraf}, F. 2003, \aap, 408, 257

\bibitem[{{Zickgraf} {et~al.}(1985){Zickgraf}, {Wolf}, {Stahl}, {Leitherer}, \&
  {Klare}}]{zickgraf1985}
{Zickgraf}, F., {Wolf}, B., {Stahl}, O., {Leitherer}, C., \& {Klare}, G. 1985,
  \aap, 143, 421

\end{thebibliography}

\end{document}